\documentclass{aa}  
\usepackage{twoopt}
\bibpunct{(}{)}{;}{a}{}{,}             
  \newcommandtwoopt{\citeads}[3][][]{\href{http://adsabs.harvard.edu/abs/#3}%
    {\def\hyper@linkstart##1##2{}%
     \let\hyper@linkend\@empty\citealp[#1][#2]{#3}}}
  \newcommandtwoopt{\citepads}[3][][]{\href{http://adsabs.harvard.edu/abs/#3}%
    {\def\hyper@linkstart##1##2{}%
     \let\hyper@linkend\@empty\citep[#1][#2]{#3}}}
  \newcommandtwoopt{\citetads}[3][][]{\href{http://adsabs.harvard.edu/abs/#3}%
    {\def\hyper@linkstart##1##2{}%
     \let\hyper@linkend\@empty\citet[#1][#2]{#3}}}
  \newcommandtwoopt{\citeyearads}[3][][]%
    {\href{http://adsabs.harvard.edu/abs/#3}
    {\def\hyper@linkstart##1##2{}%
     \let\hyper@linkend\@empty\citeyear[#1][#2]{#3}}}
\makeatother

\usepackage{graphicx}
\usepackage{footnote}
\usepackage{amsmath}
\usepackage{amssymb}
\usepackage{xcolor}
\usepackage{pdflscape}
\usepackage{url}
\usepackage[varg]{txfonts}
\usepackage{bm}		
\usepackage{booktabs}
\usepackage{txfonts}
\usepackage[pagewise]{lineno}

\newcommand{\NHI}{\ensuremath{N(\mbox{\ion{H}{i}})}}
\newcommand{\logNHI}{\ensuremath{\log N(\mbox{\ion{H}{i}})}}

\def\hi{H~{\sc i}}

\newcommand{\nhi}{\ensuremath{N(\mbox{\ion{H}{i}})}}

\def\cii{C~{\sc ii}}
\def\civ{C~{\sc iv}}

\def\mgii{Mg~{\sc ii}}

\def\ovi{O~{\sc vi}}

\DeclareGraphicsExtensions{.pdf,.png,.jpg}
\begin{document} 

   \title{MUSE-ALMA Haloes}
   \subtitle{XIV. The ALMA Large Programme Data Release}

   \author{C\'eline~P\'eroux\inst{1, 2\thanks{Email: cperoux@eso.org}}, Jianhang~Chen\inst{3}, Victoria~Bollo\inst{1}, Tamsyn O'Beirne\inst{1,4,5}, Capucine Barfety\inst{3}, Simon Weng\inst{2}, Ramona Augustin\inst{6}, Benedetta Casavecchia\inst{7}, Martin~Zwaan\inst{1}, Linda Tacconi\inst{3}, Natascha F\"oster Schreiber\inst{3}, Laurent Chemin\inst{8}, Varsha Kulkarni\inst{9}
        }

  \institute{European Southern Observatory, Karl-Schwarzschild-Str. 2, 85748 Garching near Munich, Germany \label{1}
               \and
             Aix Marseille Univ., CNRS, LAM, (Laboratoire d’Astrophysique de Marseille), UMR 7326, 13388 Marseille, France \label{2}
            \and
             Max-Planck-Institut für Extraterrestrische Physik (MPE), Giessenbachstrasse 1, 85748 Garching, Germany \label{3}
              \and
              Centre for Astrophysics and Supercomputing, Swinburne University of Technology, Hawthorn, Victoria 3122, Australia \label{4}
              \and
CSIRO Space and Astronomy, PO Box 1130, Bentley WA 6102, Australia \label{5}
              \and
              Leibniz-Institut f{\"u}r Astrophysik Potsdam (AIP), An der Sternwarte 16, 14482 Potsdam, Germany\label{6}
              \and
             Max-Planck-Institut f\"{u}r Astrophysik (MPA), Karl-Schwarzschild-Str. 1, 85741 Garching, Germany \label{7}
            \and
            Observatoire Astronomique de Strasbourg - UMR 7550, 11 rue de l'Universit\'e, 67000 Strasbourg, France \label{8}
            \and
Department of Physics and Astronomy, University of South Carolina, Columbia, SC 29208, USA \label{9}\\          }

   \date{Received XXX; accepted XXX}

\titlerunning{MUSE-ALMA Haloes ALMA LP}
\authorrunning{C.~P\'eroux et al.}
 
\abstract{
The interactions between gas and galaxies are of paramount importance to our understanding of structure formation. A
fundamental element of such baryon cycle studies is a complete census of the condensed matter (stars and cold gas) in both
galaxies and their immediate surroundings, the so-called circumgalactic medium. Of particular importance are the processes of converting molecular gas into stars and deciphering whether this $<$100K gas is tracing gas flows. The MUSE-ALMA Haloes survey has assembled probes of the neutral atomic gas \hi, ionised gas and stellar component for a sample of 79 z$\sim$0.5 galaxies. This is achieved by combining absorption line spectroscopy of background quasars { with projected impact parameter as low as 5 kpc and up to 250 kpc} to probe the low density circumgalactic gas with emission line tracers of both ionised gas and stars. This paper presents the Large Programme observational set-up and releases the data characterising the molecular phase of some of the MUSE-ALMA Haloes targets so as to reach a complete census of their condensed baryons. ALMA's unique millimetre coverage and high sensitivity is ideal to characterise the CO emission of these z$\sim$0.5 \hi-selected galaxies. By measuring the molecular mass, kinematics, gas flows and gas
fractions, this project then: i) quantifies the role of molecular gas in \hi-rich galaxies, ii) characterises the molecular phase of gas flows from morpho-kinematics analyses, and iii) establishes a census of the condensed baryons in the interstellar and circumgalactic media. Ultimately, these results will provide unique insights into the baryon cycle - a crucial component of galaxy formation.
 }

   \keywords{molecular gas -- galaxy evolution -- star formation -- ISM -- high-redshift}

   \maketitle
%
\section{Introduction} \label{sec:introduction}

The mass-energy census of the cosmos is now well constrained. The vast majority of the Universe's constituents are in the form of dark energy and dark matter which are subjects of major research. The total amount of normal baryonic matter is well measured by at least three independent techniques probing the Universe at different epochs: small temperature fluctuations in the cosmological background spectral energy distribution \citep{Planck16}, abundance of heavy elements produced during the Big Bang Nucleosynthesis \citep{cooke2014} and dispersion measures of Fast Radio Bursts \citep{Macquart20}. However, only a minority of the remaining normal baryonic matter can be probed by the observations of starlight within galaxies \citep{shull2012, nicastro2018, Mathur21, Das21}, because a major reservoir of the baryons are in the form of gas \citep{peroux2020}.  

The gas is of prime interest since it flows along the cosmic filaments and feeds the formation of galaxies, groups and clusters. Indeed, galaxy formation is fed by inflows of gas from the intergalactic medium or IGM \citep{dekel09}. Baryons from the cosmic web accrete onto galaxies, where they cool into a dense atomic and later molecular phase, which acts as the fuel for star formation. Once stars are formed, galaxies interact with the IGM by polluting it with ionising photons and chemical elements formed in stars and supernovae through galactic winds \citep{pettini2003, shull2014b, danforth16}. Understanding the drivers of the cold gas accretion rate onto galaxies, the efficiency of converting molecular gas into stars and the role of outflowing gas in preventing the infall of new accreting material, has assumed increasing importance in galaxy formation studies. 

Therefore, galaxy evolution is strongly driven by the physical processes by
which the gas flows inside and outwards of the galaxy, is converted into stars and either
ejected out through winds or recycled to form a new generation of stars. In particular, the condensed forms of this gas, notably the cold gas traced by molecular hydrogen, provide the ultimate reservoir of fuel for forming stars. This cosmic baryon cycle describes the motion and phase transformation of the baryons \citep{Tacconi20, Walter20}. These global processes take place in the immediate surrounding of galaxies, thus constituting the so-called "galaxy baryon cycle" \citep{peroux2020}.

The sites of these multi-phase gas exchanges are the immediate surroundings of galaxies, the circumgalactic medium or CGM \citep{tumlinson17, Peroux24}. With the advent of precursors of the future Square Kilometre Array Observatory (SKAO) like MeerKAT \citep{2016jonas} or ASKAP \citep{2008johnston}, surveys aim at detecting the extended reservoir of low density \hi\ emission around  galaxies in the local Universe that are likely responsible for the cold gas accretion. For instance, the MHONGOOSE \hi\ observations on MeerKAT \citep{2024deblok} is the deepest survey ever made at 21 cm, spanning $\mathrm{M_{\hi} = 10^7\, to\, 10^{11}\, M_{\odot}}$. One of the key outcomes of MHONGOOSE is the paucity of low density \hi\ emission in the atomic gas around galaxies \citep{2025marasco, 2025veronese}, in contrast with expectations from cosmological hydrodynamical simulations. Such a study demonstrates the need for even more sensitive observations with the SKAO to reach densities at least an order of magnitude deeper to search for large amounts of \hi\ gas in isolated clouds, dwarf galaxies or potential cosmic filaments of neutral hydrogen \citep{2024deblok}. 

An alternative powerful tool to study the cooler phase of that gas is provided by absorption lines in the spectra of background quasars as it enables to study regions of under-dense gas \citep{morris1993, tripp1998}. Indeed, in these quasar absorbers, the minimum gas density that is probed is set by the brightness of the background source and thus independent of the redshift of gas studied. Thus, quasar absorption lines have proven invaluable to probe the neutral atomic gas of foreground galaxies that would be otherwise too diffuse to be detected \citep{Peroux17, augustin2018}. Recently, 3D Integral Field Spectroscopy, producing data cubes where each pixel on the image has a spectrum, has proven to be a true game-changer to detect in emission the ionised gas in the circumgalactic medium of these foreground galaxies. For example, the Guaranteed Time Observations program MEGAFLOW used \mgii\ absorbers in 22 quasar fields to characterise the geometry, kinematics, and impact of gas flows around star-forming galaxies \citep{Bouche25}. A primary finding of MEGAFLOW is that the cool halo gas traced by \mgii\ absorption is not isotropically distributed around galaxies. Instead, it shows a strong bimodal distribution in the azimuthal angle relative to the galaxy's major axis \citep{Schroetter19}. By assuming that absorbers near the minor axis probe outflows \citep{Peroux20b} and making assumption on \nhi, MEGAFLOW has placed constraints on the properties of gas flows, including mass loading factor (the ratio of the mass ejection rate in the wind to the galaxy's star formation rate). MUSEQuBES is another Guaranteed Time Observations program which uses MUSE to study the CGM of galaxies \citep{Muzahid20, Dutta24}. MUSEQuBES uses other metal tracers, namely \ovi\ absorption, to study the highly ionised CGM around low-mass, dwarf galaxies at lower redshifts  \citep{Dutta24}. This gas is found to be gravitationally bound to the galaxies and the averaged \ovi\ column density around low-mass galaxies ($M_{\star}<10^{8.7}$) is significantly lower than that around more massive galaxies \citep{Dutta25a, Dutta25b}. A significant portion of the survey focuses on characterising the CGM of Ly$\alpha$ emitters at a redshift of z $\sim$ 3.3, which additionally enables estimating constraints on the \hi\ column density of the absorbers from ground-based observations. The survey found significant excess \hi\ and \civ\ absorption in the CGM of these Ly$\alpha$ emitters, extending out to at least 250 pkpc, which corresponds to about 7 times the virial radii of these galaxies \citep{muzahid2021}. A key finding is the strong influence of the galactic environment on the CGM. Ly$\alpha$ emitters which are part of pairs or groups show significantly stronger absorption and a higher gas covering fraction compared to isolated galaxies \citep{Banerjee23, Banerjee25}. The MAGG survey instead investigates the connection between optically thick gas and galaxies in the early Universe, specifically at redshifts z = 3-4 \citep{Lofthouse20, Lofthouse23, Galbiati24}. The survey has shown that Ly$\alpha$ emitters are situated within cosmic structures that contain multiphase gas, varying in composition and temperature \citep{Galbiati24}. The properties of these gas phases are strongly influenced by the large-scale galaxy environment, with the strength and kinematics of the gas correlating positively with the number of associated galaxies \citep{Galbiati24}. To benefit from the information provided by \hi\ at lower redshift, another survey dubbed CUBS uses absorption-line spectroscopy of 15 UV-bright background quasars observed with COS and WFC3 to map with MUSE the diffuse gas in and around galaxies at redshifts z $<$ 1. CUBS has shown that Lyman Limit Systems (LLSs) originate in diverse settings, including the halos of low-mass dwarf galaxy pairs, the extended gaseous disks of star-forming galaxies, the vicinity of massive quiescent galaxies, and within galaxy groups \citep{ChenHsiaoWen20}. Recently, signatures of inflows and outflows have also been reported from emission studies also using the KCWI instrument on Keck \citep{Guo23, Pessa24, Erb23}. 

Complementary to these efforts, recent years have also seen the development of similar approaches at mm wavelengths. Despite its modest primary beam, ALMA can be seen as a 3D spectrograph since it provides simultaneously frequency and spatial information. On top of the early efforts of the MUSE-ALMA Haloes survey itself \citep{Klitsch18, Peroux19}, \cite{Kanekar18} focus on the nature of galaxies that give rise to high-metallicity damped Lyman-$\alpha$ (DLA) absorption systems at intermediate redshifts (z = 0.5–0.8) by searching for carbon monoxide (CO) emission, a tracer of molecular gas, from seven of these absorption-selected galaxies. The authors report high molecular gas fractions, the ratio of molecular gas mass to the total baryonic mass, and low SFR in the five systems they detect. Interestingly, \cite{Klitsch21} report a high detection rate in the detection of CO emission in the fields of 6 H$_2$-bearing absorbers. Again, they find that the majority of the H$_2$ absorbers are associated with multiple galaxies rather than single, isolated galactic halos. Recently, \cite{Neeleman25} presented the results of a survey studying galaxies associated with damped Ly$\alpha$ absorbers (DLAs) at a redshift around 4. The study found that the ability to detect the host galaxies of these DLAs in [\cii] emission is strongly dependent on the metallicity of the absorbing gas itself. The detection rate is high (71\%) for DLAs with a metallicity [M/H] $> -$1.5, but zero for those with lower metallicities. Otherwise, they find that the detected galaxies have properties similar to typical star-forming galaxies at this high redshift.

By combining 3D information from both optical and millimetre wavelengths, the MUSE-ALMA Haloes survey builds on both these lines of effort. With the additional high-spatial resolution imaging enabled by space missions, the stellar continuum for the same objects is now robustly traced despite the light contamination from the bright background quasar \citep{augustin2018, Karki23, Augustin24, Karki26}. Without a quantitative description of the molecular phase however, one cannot reach a full understanding of the baryon cycle. 

In this context, this paper introduces the ALMA Large Programme component of the MUSE-ALMA survey whose goal is indeed to probe the molecular phase of an otherwise well-studied sample of z$\sim$0.5 galaxies. The overarching objective of the MUSE-ALMA Haloes project is to use multi-wavelength information to probe the physical processes by which gas is transformed into stars through the baryon cycle. This paper is organised as follows. 
In Section~\ref{sec:target_selection}, we describe the selection of the MUSE-ALMA Haloes targets.
Section~\ref{sec:MAH} presents the multi-wavelength complementary datasets and the physical properties they constrain. 
Section~\ref{sec:ALMA_obs} describes the ALMA observations, while Section~\ref{sec:LP} specifically details ALMA Large Programme survey design, describes the observational set-up and an extensively reports on the data reduction processes. 
Section~\ref{sec:products} introduces the release of the resulting data products, while Section~\ref{sec:science} summarises the science goals of the programme. Finally, Section~\ref{sec:conclusions} provides our key conclusions. 
Throughout this paper, we use $H_0 = 70$ kms$^{-1}$Mpc$^{-1}$, $\Omega_{\text{M}} =0.3$ and $\Omega_{\Lambda} = 0.7$.

\section{The MUSE-ALMA Haloes primary target selection} 
\label{sec:target_selection}

The MUSE-ALMA Haloes survey is a multi-facility programme conducted through various distinct observing campaigns over the years. Several similar efforts have come to fruition recently. This includes the impressive PHANGS which is a large, multi-wavelength observational program designed to build a comprehensive dataset for studying star formation and the interstellar medium (ISM) across a representative sample of nearby spiral galaxies \citep{Schinnerer19, LangPhilipp20, Leroy21}. By combining data from ALMA, MUSE, HST, and now JWST, PHANGS enables a multi-phase, multi-scale investigation of the entire star formation cycle \citep{Barnes23, LeeJanice23}. Likewise, MAUVE is a MUSE and ALMA multiwavelength program designed to investigate how the cluster environment shapes the gas–star formation cycle in galaxies falling into the Virgo Cluster \citep{Watts24}. MAUVE aims to characterise the interstellar medium, stellar populations, and star formation activity in 40 late-type Virgo Cluster galaxies at various stages of infall. The upcoming KILOGAS survey which combines maps at 1 kpc resolution in a sample of over 400 galaxies with available optical integral field spectroscopy from the MaNGA and SAMI surveys, provides resolution-matched SFR, metallicity, and stellar maps, as well as ionised gas and stellar kinematics. In the future, with the availability of new channels across observatories\footnote{\url{www.eso.org/sci/observing/phase1/JointVLT-ALMA.html}}, such multi-wavelength approach is likely to be key to unveil multi-phase physics in various galactic media. 

The MUSE-ALMA Haloes survey uniquely provides accurate information on the \hi\ content and metallicity of the CGM of its targets. Because of the asynchronous nature of the proposing cycles, the target selection of the MUSE-ALMA Haloes sample has been evolving as new information has become available. 

\begin{figure}[h]
\centering
    \includegraphics[width=7.5cm, angle=0]{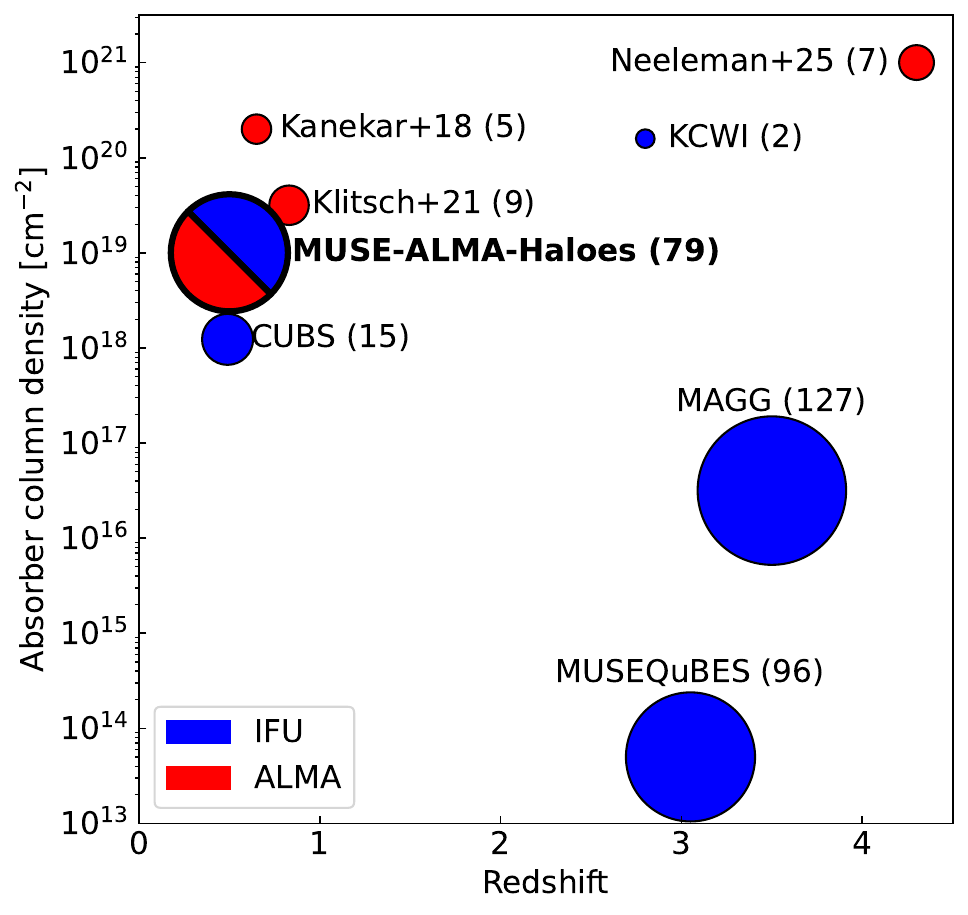}
    \caption{{MUSE-ALMA Haloes surveys in the context of other comparable efforts.} This figure reports surveys targeting quasar absorbers with known \nhi\ column densities as a function of redshift \citep[this work,][]{Muzahid20, ChenHsiaoWen20, Lofthouse23, Erb23, Kanekar18, Klitsch21, Neeleman25}. "IFU" refers to both VLT/MUSE and Keck/KCWI. Symbol sizes are proportional to the number of targets in each surveys. On top of the sizeable sample, the MUSE-ALMA Haloes survey uniquely combines 3D observations at optical and millimetre wavelengths to enable studies of the multi-phase gas.  }
\label{fig:other_surveys}
\end{figure}

In a nutshell, the quasar absorbers of the MUSE-ALMA Haloes survey were selected on HST/UV spectroscopy to have the \hi\ column density, log [N(H\,I)/cm$^{-2}$]$>$18 and $z_{\rm abs}<0.85$. The first of these two elements of the selection is key to enable robust absorption metallicity measurements \citep{Halley26} and gas mass estimates in galactic flows \citep{Rahmani18b}. Indeed, we emphasize that knowing accurately the \hi\ column density in absorption is key to measure precisely the neutral gas metallicity and assess the potential gas ionisation correction \citep{Halley26}. In addition, \nhi\ is a key component of the estimate of the mass loading factor \citep{bouche2013, schroetter2015}. This selection criterion adds to other surveys focusing on metal lines \citep{Bouche25, Dutta25a}. 

On the other hand, the redshift criterion of the target selection enables both accurate redshift measurements and kinematic reconstruction \citep{Szakacs21, Weng23b}. This redshift range enables the robust determination of both the systemic redshift based on narrow emission lines (as opposed to the resonant Ly$\alpha$ line), as well as the star formation rates and emission metallicity of these galaxies based on rest-frame optical emission diagnostics. Furthermore, at low-redshift, the larger apparent size and increased surface brightness enable kinematic reconstruction of the objects. Finally, the redshift range set to be $z_{\rm abs}<0.85$ complements other efforts at $z>3$ \citep{muzahid2021}.  

At the start of the survey, the sample resulting from this selection comprises 32 individual absorbers in 19 distinct quasar fields (i.e. multiple absorbers per quasar spectra) which are dubbed "primary targets". Figure~\ref{fig:other_surveys} presents this number of targets in context of similar surveys targeting quasar absorbers with known \nhi\ column densities as a function of redshift \citep{Muzahid20, ChenHsiaoWen20, Lofthouse23, Erb23, Kanekar18, Klitsch18, Neeleman25}. On top of the sizeable sample, the MUSE-ALMA Haloes survey uniquely combines 3D observations at optical and mm-wavelengths to enable studies of the multi-phase gas. 

\begin{figure*}
\centering
	\includegraphics[width=5.3cm]{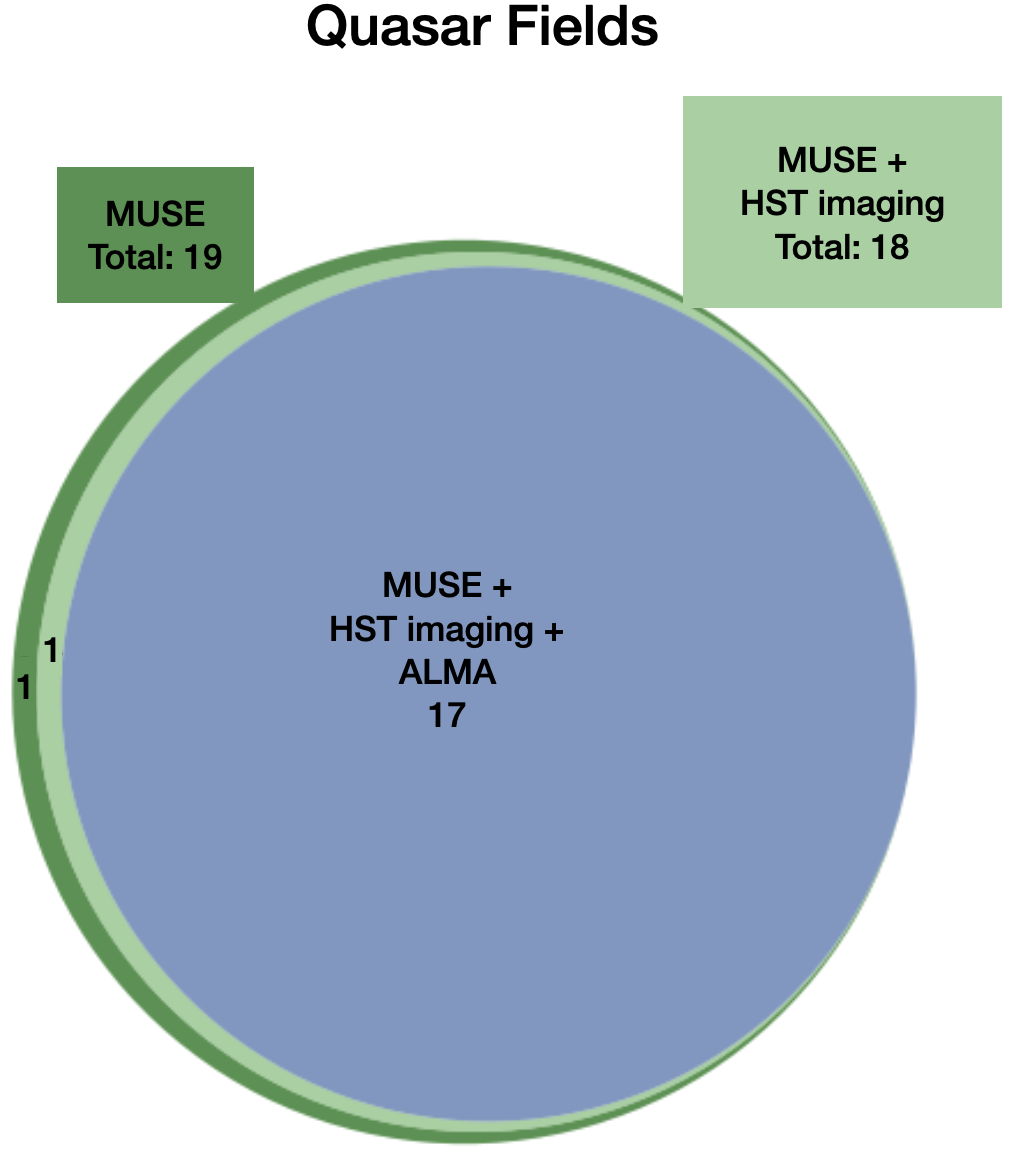}
    \includegraphics[width=5.3cm]{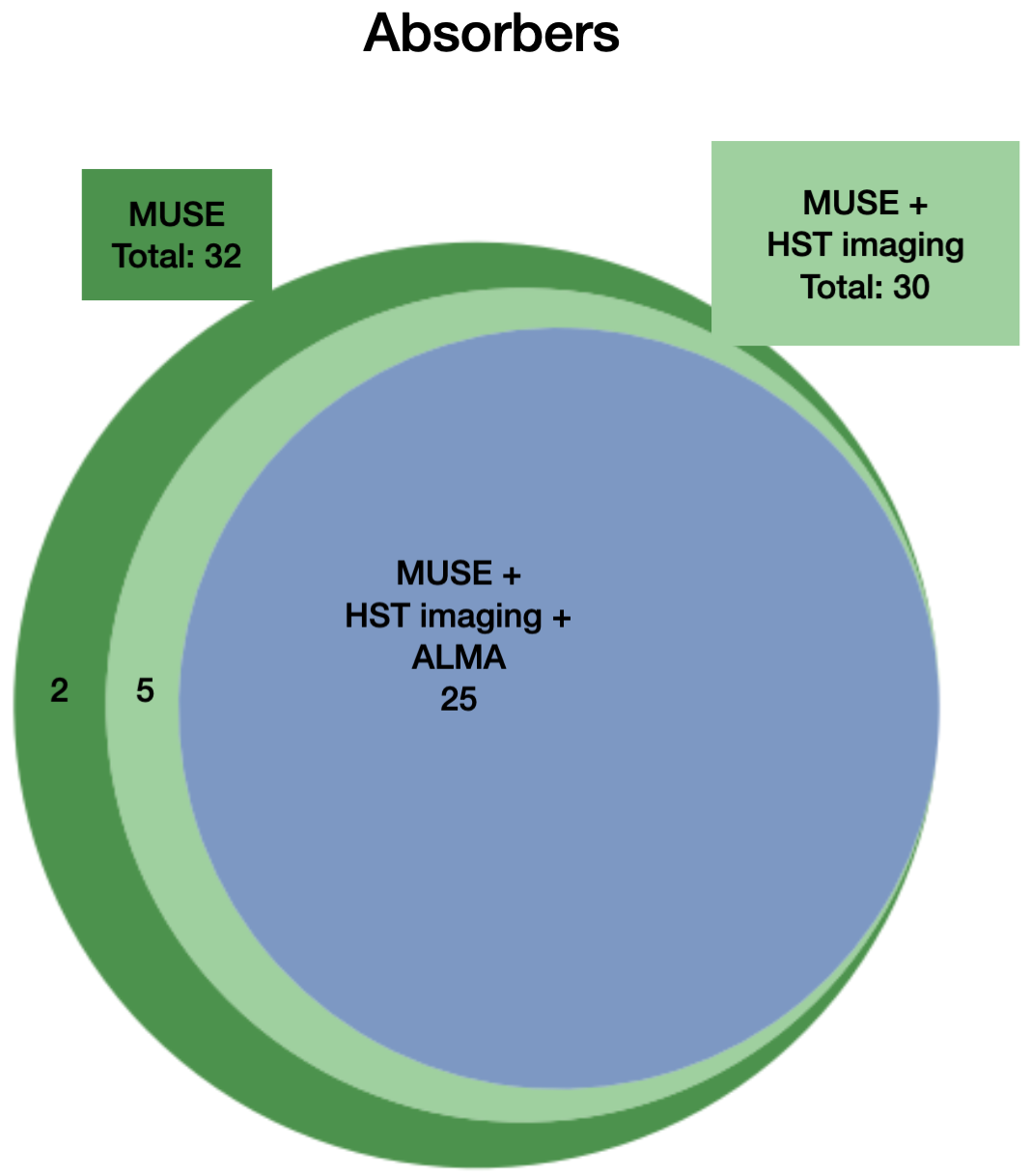}
    \includegraphics[width=5.1cm]{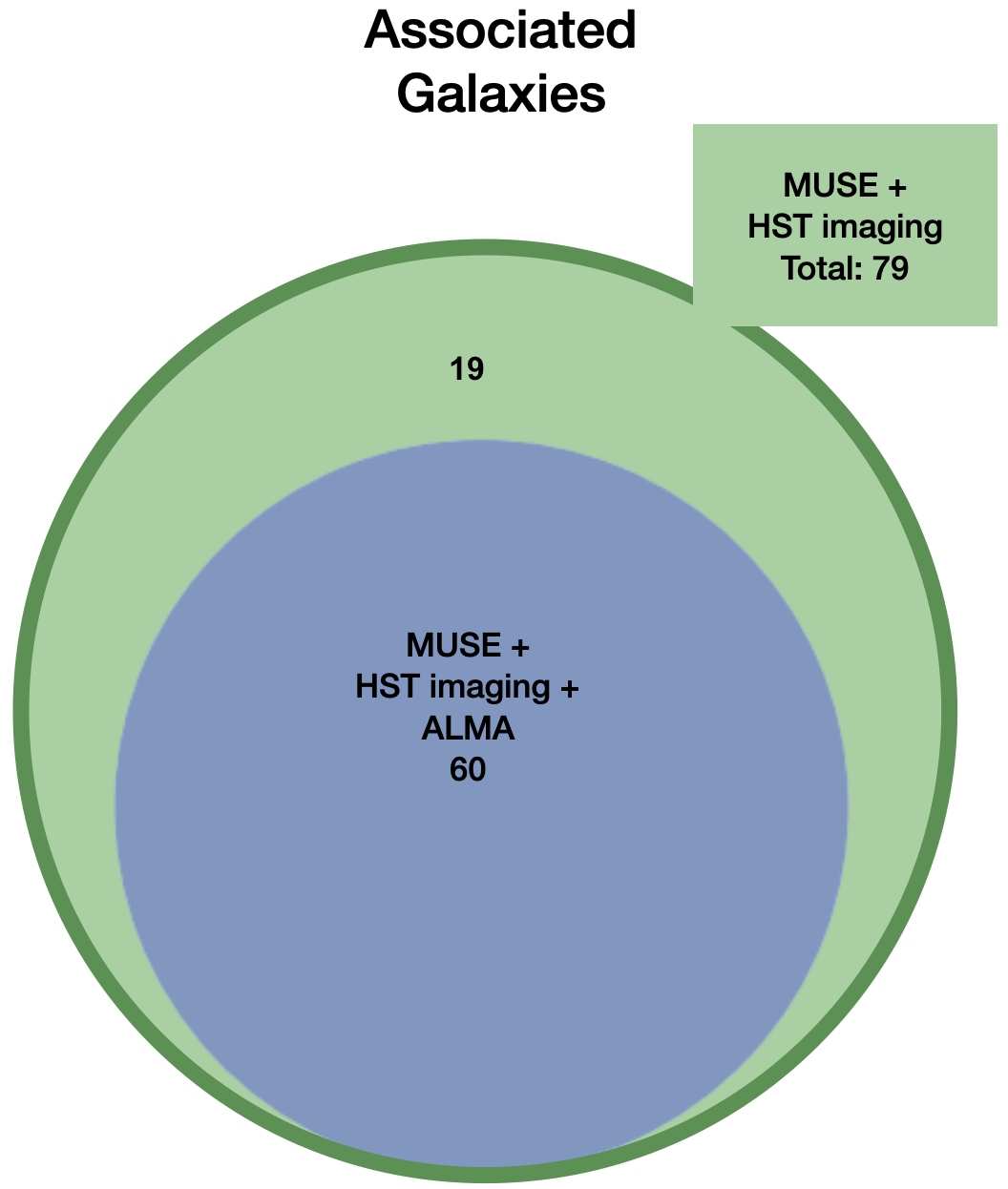}
    \caption{{MUSE-ALMA Haloes number of targets covered by various observing campaigns.} This figure provides a  graphical viewing of the numbers listed in Table~\ref{tab:MAH_sample}. The left, middle and right panels illustrate the number of quasar fields, quasar absorbers and galaxies associated with the quasar absorbers observed by each facility.}
    \label{fig:Venn_diag}    
\end{figure*}

\begin{table}
\begin{center}
\caption{{ MUSE-ALMA Haloes number of targets covered by various observing campaigns}.}
\begin{tabular}{lccc}
\hline\hline
Facility &\# quasar  	  &\# absorbers &\# associated \\
&fields &&galaxies\\
\hline                                                        
HST spectroscopy    &19 &32 &--\\
UVES/HIRES spec    &18 &31 &--\\
VLT/MUSE                &19 &32 &79\\
HST imaging         &18 &30 &79\\ 
ALMA                &17 &25$^a$ &60$^b$\\ 
\hline\hline 				       		         	 
\label{tab:MAH_sample}                                         
\end{tabular}			       			 	 
\begin{minipage}{90mm}
Note: The table summarises the number of quasar fields, total number of absorbers in all quasar fields, and galaxies associated with the absorbers in all quasar fields observed by each facility. The quasar fields were initially selected from HST UV-spectroscopy, with additional or followed-up observations from the ground with either Keck or VLT. The associated galaxies were identified with MUSE and HST imaging, while to optimise the survey strategy given the size of the primary beams, ALMA observations targeted a subset of the known galaxies. \\
$^a$ 17 (out of these 25) absorbers were observed as part of the ALMA Large Programme.\\
$^b$ 39 (out of these 60) associated galaxies were observed as part of the ALMA Large Programme.
\end{minipage}
\end{center}			       			 	 
\end{table}

Archival and new proprietary observations with VLT and Keck provide additional optical spectroscopy of the background quasars. All the quasar spectra are observed at high-spectral resolution but one (Q1110$+$0048 which contains one absorber) due to the lack of completion of the UVES PI-run. This makes a sample of 18 quasar fields, with 31 absorbers.

All 19 fields are followed-up with VLT/MUSE to detect in emission the galaxies associated with the absorbers { with impact parameters ranging from b=0 kpc for emission-line objects \citep[thanks to spectral-PSF subtraction, see][]{Peroux19} to $\sim$250 kpc for all galaxies}. Seventy-nine such emitters are detected { within $\pm$ 500 km/s of the absorber redshift} \citep{Weng23a}. An HST medium broad-band imaging program characterised the stellar content of all these objects but for one quasar field (Q0058$+$0019 with 2 absorbers but no galaxies identified in the VLT/MUSE observations). This leads to a sample of 18 quasar fields, with 30 absorbers and the full 79 associated galaxies. The ALMA observations spread over several cycles to cover all these fields but one (Q2353$-$0028 which is found to have no associated galaxies within 500 km/s in the MUSE observations). Because of the restricted primary beam however, not all the known associated galaxies detected in MUSE are targeted with ALMA. The final sample thus contains 17 quasar fields, together containing 25 absorbers. These absorbers are associated with 60 distinct galaxies. Thirty-nine of these galaxies have been targeted by the ALMA Large Programme presented in this paper. These samples are listed in Table~\ref{tab:MAH_sample} and illustrated by the Venn diagrams displayed in Figure~\ref{fig:Venn_diag}. In addition to the "primary targets", a total of 3658 galaxies are detected in the MUSE and HST datasets and are the subject of parallel studies \citep{Peroux22}.

\section{The MUSE-ALMA Haloes multi-facilities datasets} 
\label{sec:MAH}

\begin{figure*}[h]
\sidecaption
\includegraphics[width=12.5cm, angle=0]{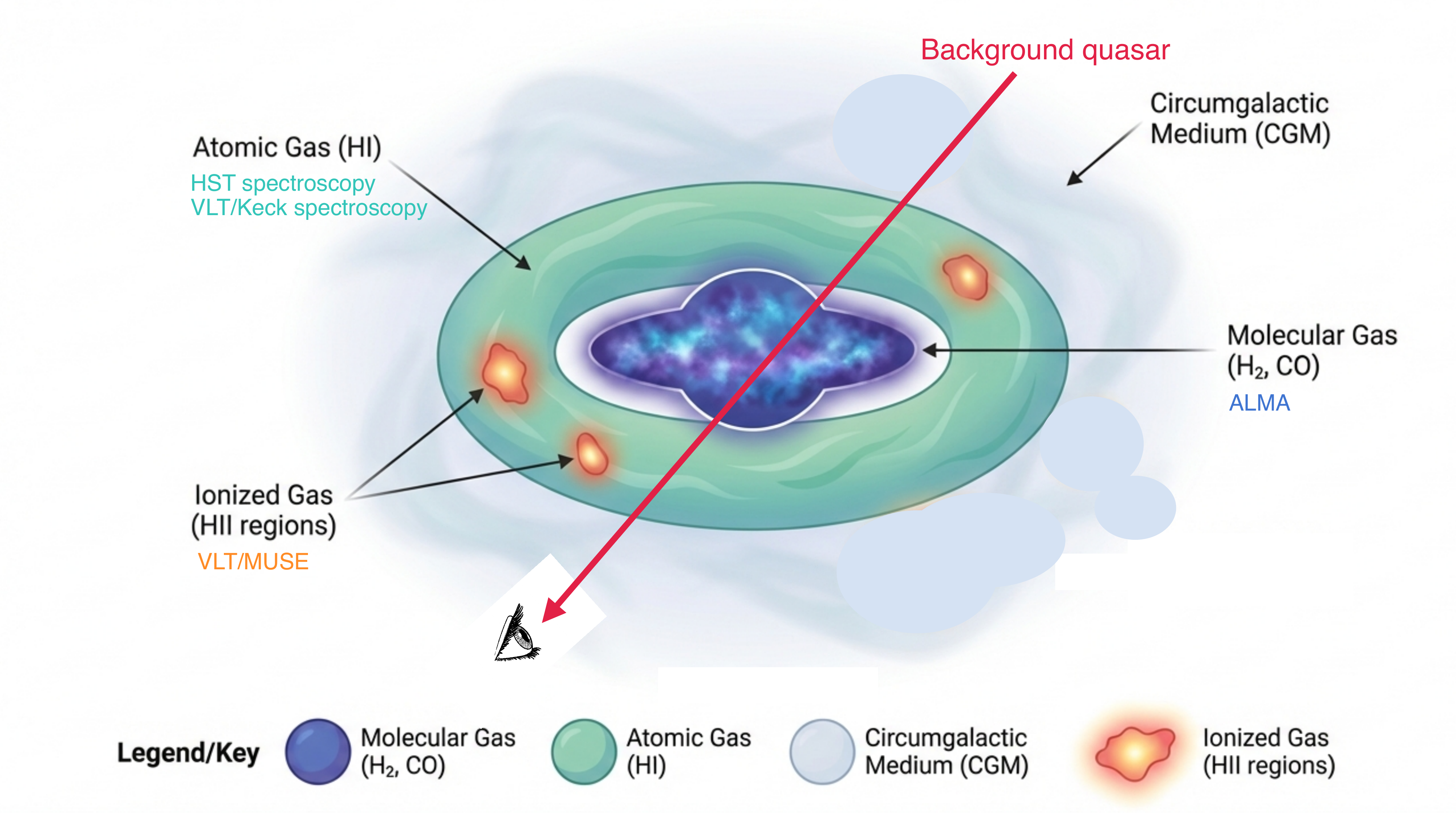}
  \caption {{Sketch of multi-phase circumgalactic and interstellar media.} The MUSE-ALMA Haloes uniquely probes the various component of the condensed baryonic matter as indicated in the labels: neutral molecular \& atomic and ionised gas and stellar components of galaxies. This is done through various observing programmes gathering data from multi-frequency facilities as indicated in colours. }
\label{fig:sketch}
\end{figure*}

In the MUSE-ALMA Haloes survey, several distinct observational campaigns provide information on the multi-phase interstellar and circumgalactic media of the galaxy sample. Indeed, the project combines absorption and emission data probing the atomic, molecular and ionised gas as well as stellar populations in these z$\sim$0.5 objects. This section reviews these datasets complementary to the ALMA observations and the physical properties they constrain. { Table~\ref{tab:publi} in Appendix lists the MUSE-ALMA Haloes survey publications to date.} The overall approach is schematically illustrated by Figure~\ref{fig:sketch}, a sketch partially AI-generated with \textsc{FigureLabs}\footnote{generated on 12$^{\rm th}$ December 2025 with \url{https://chat.figurelabs.ai/} }.

\subsection{Atomic gas}

\subsubsection{Neutral hydrogen gas from space-based UV spectroscopy}

Each background quasar has archival HST/UV spectroscopy providing reliable N(H\,I) measurements of the strong intervening low-redshift absorbers with log [N(H\,I)/cm$^{-2}$]$>$18. The observations were undertaken with either HST/FOS \citep{Harms91}, COS \citep{Green12} or STIS \citep{Kimble98}. The \hi\ column density measurements quantifies the neutral phase of the circumgalactic gas at an impact parameter of the associated galaxies and provides accurate absorption redshift measurements. For lower columns densities, where the gas could be partly ionised, photo-ionisation models with the \textsc{CLOUDY} software enabled robust estimates of the ionisation fraction of the gas \citep{Halley26}. The resulting quantities are provided in columns 2 and 3 of Table~\ref{tab:abs} and their distributions are plotted in the left and central panels of Figure~\ref{fig:abs_histo}.

\subsubsection{Gas-phase metallicity from ground-based optical spectroscopy }

Complementary to space-based HST/UV spectroscopy, ground-based spectroscopy of the background quasars provides additional information on the abundance, element ratio and dust properties of the neutral circumgalactic gas. A number of existing spectra are available through the VLT and Keck archives as displayed in the last column of Table 1 of \cite{Peroux22}. Recently, most of the remaining quasars were observed with VLT/UVES spectroscopy under a new program (ID: 113.A-0369; PI: C. Peroux). 

\begin{figure*}
\centering
	\includegraphics[width=20.5cm]{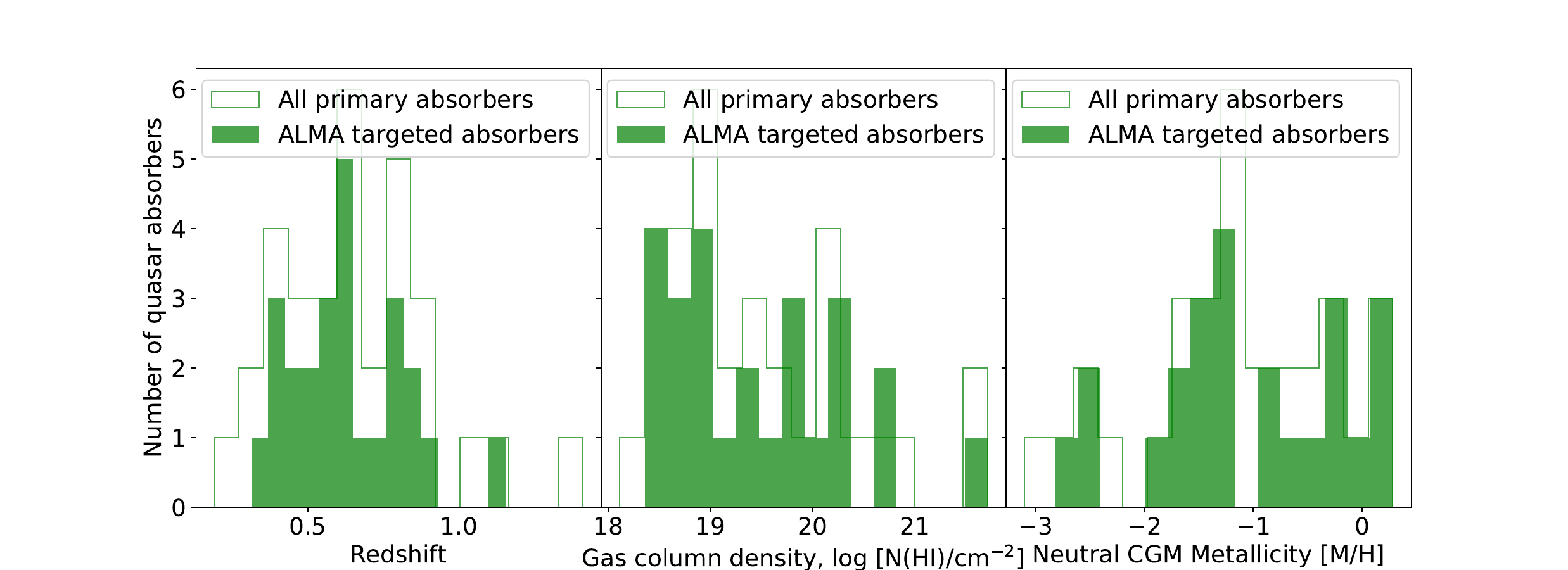}
    \caption{{Properties of the "primary targets" observed with ALMA.} {\it Left panel:} Redshift distribution of the 32 quasar absorbers making up the primary sample of MUSE-ALMA Haloes survey. The filled histogram (green) displays the redshift distribution of the 25 intervening absorbers targeted by ALMA observations. {\it Middle panel:} Atomic hydrogen gas column density distribution of these strong absorbers. The selection criterion imposes the cut at log [N(H\,I)/cm$^{-2}$]$>$18. {\it Right panel:} Neutral gas absorption metallicity measured from archival and new VLT and Keck observations after ionisation and dust correction \cite{Halley26}. The metallicity is the abundance of all metals, is expressed as $[{\rm M/H}] = \rm \log \{ N({\rm M})/N({\rm H}) \} - \log \{ N({\rm M})/N({\rm H}) \}_{\odot}$. }
    \label{fig:abs_histo}
\end{figure*}

When available, archival Keck/HIRES, VLT/UVES and VLT/X-Shooter data are used. The high-resolution quasar spectra are used to search for new quasar absorbers beyond the "primary targets". Sixteen such systems are discovered so that a total of 48 absorbers with \nhi\ measured are now available from the MUSE-ALMA Haloes survey. These will be presented in a forthcoming paper \cite{Halley26}. In addition, the quasar spectra were used to measure the metal line column densities of ions including FeII, SiII, SiIII, SiIV, dust-free ZnII, CIV from Voigt profile fitting and performing ionisation and dust depletion corrections where needed \citep{Halley26}. These high-resolution data characterise the neutral circumgalactic gas metallicity. Early results from the archival data \citep{Weng23a} indicate that the metallicity of associated galaxies is typically larger than the absorber metallicity, which decreases at larger impact parameters. The resulting absorption metallicities are provided in columns 4 of Table~\ref{tab:abs} 
and their distribution is plotted in the right panel of Figure~\ref{fig:abs_histo}.

\subsection{Galaxy emission properties from optical 3D spectroscopy}

3D Integral Field Units (IFU), which produces data cubes where each pixel on the image has a spectrum, have provided a powerful technique to associate quasar absorbers with galaxies. Building on early efforts with near-infrared IFU VLT/SINFONI \citep{Bouche07, Peroux11a, Peroux11b, Peroux13}, the potential of this technique for studying the CGM of intervening galaxies has grown with the availability of large optical IFU such as VLT/MUSE and Keck/KCWI \citep{Bouche16, Peroux17, Zabl19, Lofthouse20, Nielsen20}. Over several observing periods, the MUSE-ALMA Haloes survey has used the VLT/MUSE in particular to identify 79 galaxies responsible for the strong z$\sim$0.5 quasar absorbers making up the "primary targets". 

\begin{figure*}
\centering
\includegraphics[width=20.5cm]{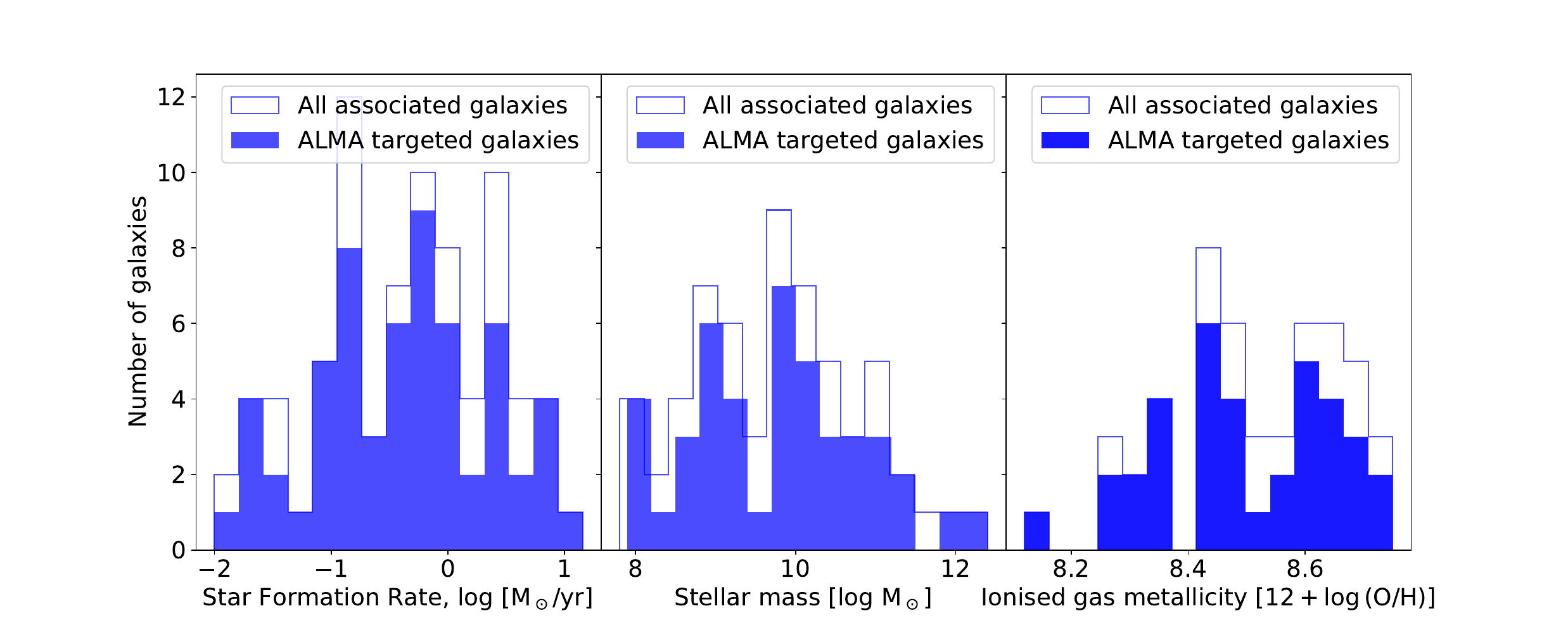}
    \caption{{Properties of the associated galaxies.} {\it Left panel:} Star Formation Rate distribution of the 79 galaxies making up the MUSE-ALMA Haloes survey. The filled histogram displays the SFR of galaxies observed with ALMA. {\it Middle panel:} Stellar mass distribution {\it Right panel:} Ionised gas emission metallicity. }
    \label{fig:gal_histo}
\end{figure*}

By detecting the galaxies in emission, it is possible to connect the circumgalactic gas seen in absorption with ionised gas detected in emission. Indeed, the lower-density extended circumgalactic gas seen in absorption can then be connected with the higher-density ionised gas detected in emission and likely located closer to/in the galaxies. \cite{Weng23a} use the measurements of the redshifts and impact parameters (provided in columns 5 and 6 of Table~\ref{tab:abs}) with respect to the background quasars of the 79 detected galaxies to statistical map the neutral circumgalactic gas around galaxies. This work also reports \nhi\ column density of the absorbers is anticorrelated with the impact parameter (scaled by virial radius) of the nearest galaxy as expected from simulations. \cite{Weng23a} additionally provide accurate estimates of the SFR and emission metallicities of the objects as reported in column 7 and 9 of Table~\ref{tab:abs} in the appendix and plotted on the right and left panels of Figure~\ref{fig:gal_histo}. Furthermore, \cite{Weng23b} make use of the 3D spectroscopy to reconstruct the kinematics of the ionised gas and establish a connection with the circumgalactic gas probed in absorption. Together these works report that 10\% of absorbers are consistent with gas accretion, a third trace outflows while the reminding systems are associated with gas in the galaxy disk, the intragroup medium or possibly low-mass galaxies below the MUSE detection limit. For comparison, the MEGAFLOW survey contains 79 strong MgII absorbers with 86 associated galaxies. Assuming that absorbers near the minor axis probe outflows, \cite{Schroetter19} report 26 galaxy–quasar pairs suitable for studying winds (i.e. 30\%). Keck/KCWI has also been used to study targetted galaxies with outflows \citep{ChenYuguang21, Erb23}. IFUs thus provides a key element of probing the baryon cycle in the MUSE-ALMA Haloes galaxies.

\subsection{Stellar properties from space-based broad-band imaging}

An essential component of the MUSE-ALMA Haloes project is a Cycle 27 forty-orbit medium Hubble Space Telescope programme (PI: C. Peroux) providing broad-band imaging with WFC3 on
both the optical (UVIS) and infrared (IR) detectors for 18 quasar fields. We also use an additional 27 orbits of archival data recorded with WFPC2, making a total of 67 orbits dedicated to this part
of the project. The properties of the objects (catalogues of sky positions, magnitudes in different bands) have been compiled and made publicly available \citep{Peroux22}. We have performed a detailed analysis of these datasets, providing measurements of the stellar mass, dust content, age of the stellar population, as well as new constraints on their resolved morphology and large-scale environment. The observations also enabled the determination of the stellar masses
of the associated galaxies through spectral energy distribution fits  \citep{Augustin24} and detailed morphological analysis \citep{Karki23}. 

\cite{Karki23} find that absorption-selected galaxies are consistent with the star formation main sequence of the global population. This work combines multi-wavelength information to demonstrate that more massive absorber counterparts are more centrally concentrated and are larger in size. \cite{Karki23} further report that SFR surface densities of absorption-selected galaxies are higher than predicted by the Kennicutt–Schmidt relation for local galaxies, suggesting a higher star formation efficiency in the \hi-rich galaxies. \cite{Augustin24} establish that these objects probe a wide range of stellar masses: 8.1 $<$ log($M_*$/M$_{\odot}$) $<$ 11.8, as illustrated in the middle panle of Figure~\ref{fig:gal_histo}. The galaxy Q0454$-$220$\_$4 stellar mass has been recently revised to be M$_*=10^{\rm 10.6\pm0.1}$  M$_{\odot}$. Given their star-formation rates, most of these objects lie on the main sequence of galaxies. 

\cite{Augustin24} also confirm a previously reported anti-correlation between the stellar masses and CGM hydrogen
column density \nhi, indicating an evolutionary trend where higher mass galaxies are less likely to host large amounts of neutral atomic \hi\ gas in their immediate vicinity up to 120 kpc. Recently, \cite{Karki26} made use of HST deep imaging and MUSE 3D spectroscopy combined to study the low-surface brightness extended emission around some of the galaxies associated with quasar absorbers. The authors find that galaxies exhibiting these features have enhanced star formation rates, specific star formation rates, SFR surface densities, emission metallicities, and dust reddening compared to galaxies without extended features. These findings point at signatures of galaxy interactions and possible tidal streams. \cite{Karki26} also note that features have higher sSFR and higher ionisation parameters but lower SFR, lower emission metallicities, and lower dust reddening than their host galaxies indicating a different physical origin of the media.

\subsection{Molecular gas content from ALMA observations}

ALMA's mm coverage and high sensitivity has proven powerful to characterise the molecular emission of intervening galaxies. Over the last decade, CO at z$<$4 and [CII] at higher redshift have been used to detect in emission galaxies associated with strong N(H\,I)-quasar absorbers. \cite{Neeleman16, neeleman2018, augustin2018, Moller18, Fynbo18, Neeleman20, Kanekar20} and others report the ALMA observations of quasar absorbers in the range z=1--3. At the same time, the MUSE-ALMA Haloes survey undertook several ALMA runs to probe and analyse the molecular content of the associated galaxies. While the following section~\ref{sec:ALMA_obs} will provide technical details on each of the ALMA datasets, this section summarises findings from early studies.

\cite{Klitsch18} used multiple CO transitions and mm continuum detections to infer that the \nhi\ absorber in the MUSE-ALMA Haloes Q0420$-$0127 quasar field traces either an outflow from the most massive galaxy or intragroup gas. \cite{Peroux19} report 3 CO detections associated with another MUSE-ALMA Haloes quasar absorber, finding long depletion time-scales for the molecular gas and kinematics consistent with the ionised gas. \cite{Szakacs21} present observations of another 6 absorbers associated with 9 galaxies, with multiple galaxies in overdensity being related to the absorbers \citep[see also][]{Klitsch21}. \cite{Szakacs21} perform multi-phase kinematic studies which show that the molecular and ionised gas phases are well aligned directionally and that the maximum rotation velocities closely match. Their results demonstrate that the two gas phases within the disk are strongly coupled. The authors also report evidence for infalling gas. Together, these studies evidence the power of combining multi-wavelength datasets to gain important information on the cool baryons in a bounded structures at intermediate redshift.

\subsection{Astrometry across multi-wavelength datasets} 
\label{sec:Astrometry}

A key component of the MUSE-ALMA Haloes survey is the multi-frequency observational imaging campaign conducted with VLT/MUSE, ALMA and HST. In order to intercompare these various datasets, it is essential for the observations to be recorded on a common astrometric frame. Specifically, detailed studies of the kinematics of multi-phase gas traced by MUSE and ALMA require accurate alignment \citep{Peroux19, Szakacs21}. Likewise, search for CO emission in ALMA data based on MUSE and/or HST galaxy spatial position entailed precise astrometry \citep{augustin2018, Bollo26}. For all these reasons, we take the opportunity of the ALMA Large Programme being completed to make a reappraisal of the astrometry across the various datasets presented in Section~\ref{sec:MAH}. 

The absolute position accuracy of an ALMA image is theoretically limited by the accuracy of the phase calibrator position and the quality of the phase transfer during the observations. The latter depends on the weather conditions, the angle between the phase calibrator and the centre of the targeted field, the cadence of observations of the phase calibrator, and the level of errors in the antenna position measurements. For observations of the type of the ALMA MUSE Haloes survey, the position accuracy is less than the synthesized beam width tabulated and greater than 0.1 arcseconds. Therefore, while the primary beam of the ALMA observations is small with respect to the MUSE and HST images, the accuracy of the astrometry is the most suitable for all datasets for our scientific purposes.

On the other hand, the HST broad-band images at hand \citep{Karki23} provide large fields-of-view. As a result, a number of bright objects, typically stars, are covered by each of the space telescope pointings and in every filters. Based on previous experience \citep{Peroux19}, we expect some of these images to be slightly offset but also small rotation of the fields around the central pointing are not unusual. Thanksfully, the European Space Agency Gaia with its average high positional accuracy of 0.05 milli-arcsecond provides an invaluable reference to evaluate the fidelity of the HST images. The Gaia Early Data Release 3 (EDR3) which took place in December 2020 has a significant astrometric (positions, parallaxes, proper motions) and photometric (integrated G, GBP, GRP) improvements \citep{Gaia21}. To solve for accurate and reliable astrometry, we use the positions of several stars detected both in HST images and present in the Gaia catalogue. Whenever possible, we use 3 bright objects to solve for potential rotation of the HST fields. The resulting offsets are sub-arcsec with a mean value of 0.235 arcsec. After correction, we find the alignment of our targets between the HST and ALMA data to be excellent.

Finally, the 1-arcmin squared MUSE field-of-view also often includes multiple bright objects. The MUSE individual sub-exposures were aligned with the ESO pipeline's \texttt{exp$\_$align} recipe using these point sources resulting in an accurate relative astrometry in the combined single data cube \citep{Peroux22}. To reach absolute astrometry, we now use the bright stars to calculate the offset with respect to the corrected HST broad-band images. The systemic offsets are in most cases smaller than the beam size ($\sim$0.8 arcsec). In order to provide an uniform dataset for various science projects across the collaboration, we choose to edit the HST images and MUSE cubes header files to correct to the Gaia DR3 astrometry. As a result, the multi-wavelength observations of the MUSE-ALMA Haloes survey are consistently aligned to the absolute astrometry newly provided by ALMA and Gaia DR3.


\section{The MUSE-ALMA Haloes ALMA observations} 
\label{sec:ALMA_obs}

\subsection{Primary targets from pilot programmes}

Before embarking into a Large Programme, the MUSE-ALMA Haloes survey demonstrated feasibility of the observing strategy by targeting a few selected "primary targets" in early Cycles. The results of these observations have been presented in several publications, including \cite{Peroux19, Klitsch18, Szakacs21}. Specifically, the quasar field Q1130$-$1449 was observed in Cycle 4 (PI: C. Peroux), while 9 more associated galaxies were part of a Cycle 5 programme (PI: C. Peroux). Additionally, a Cycle 6 run (PI: A. Klitsch) targeted one more quasar field (Q0420$-$0127). These early observations informed the optimal strategy for the subsequent Large Programme.

\subsection{Primary targets from ALMACAL}

We further make use of ALMACAL to search for overlap with potential MUSE-ALMA Haloes "primary targets". ALMACAL\footnote{\url{https://almacal.wordpress.com/}} is a database of reprocessed ALMA phase and amplitude calibration scans \citep{Bollo24}. In its latest release (ALMACAL-22), the data accumulated from the start of ALMA operations until May 2022 make up a total of over 2000 hours of integration time over a large area and at all frequencies. ALMACAL has already enabled a number of specific science findings including measurements of the molecular gas mass density cosmological evolution free of cosmic variance \citep{Klitsch19b, hamanowicz2020, Bollo25} as well as complete mm number counts \citep{ChenJianhang23} and other scientific results \citep{Klitsch19a, Klitsch20}. The cross-match search between ALMACAL and MUSE-ALMA Haloes results in one target (Q0420$-$0127) with sufficient observations to enable detailed study of the system \citep{Klitsch18}, while more valuable observations might become available in the future as part of the ever-growing ALMACAL database.

\subsection{Primary targets from ALMA archives}

We also cross-match the ALMA public archives\footnote{\url{https://almascience.eso.org/aq/}} with MUSE-ALMA Haloes "primary targets". This search led to multiple matches as reported in Table 3 of \cite{Peroux22}. In particular, two of the MUSE-ALMA quasar fields (Q0058$+$0019 and Q0138$-$0005) have been observed with sufficient depth as part of Cycle 1 (PI: X. Prochaska) and Cycle 3 (PI: N. Kanekar). \cite{Kanekar18} report results from these studies which are included in the following analysis as part of the so-called previous ALMA observations.

\subsection{Primary targets from ALMA large programme}

In Cycle 10, we were awarded a Large Programme (PI: C. Peroux, V. Kulkarni, L. Chemin) to complement these early observations in an efficient way and provide information on the molecular content of most of the MUSE-ALMA Haloes survey "primary targets". The programme targeted 39 distinct galaxies known to be associated with strong \hi\ quasar absorbers identified with MUSE and HST. The design of the survey was made efficient by aiming at several objects per pointings, also including mosaicing of some of the fields. \cite{Bollo26} report on the detections of these strong \nhi\ quasar absorbers and their associated physical properties.

\subsection{Serendipitous detections from ALMA large programme} 

In addition to the "primary targets" which correspond to MUSE-detected galaxies associated with strong \nhi-absorbers, the ALMA Large Programme cubes contain a number of additional emission line detections also augmented with multi-wavelength observations. While these are not
directly the targets of the primary sample, some of the science described
in Section~\ref{sec:science} is also addressed by these serendipitous detections.
These typically include CO emitters at intermediate redshifts, some of which are extremely mm-bright, and are the focus of further studies \citep{Klitsch20, ChenJianhang23}. We refer to this sample as the "serendipitous detections".

\section{ALMA large programme}
\label{sec:LP}

\subsection{Survey design}
\label{sec:design}

\begin{figure*}[h]
\centering
\includegraphics[width=16.5cm, angle=0]{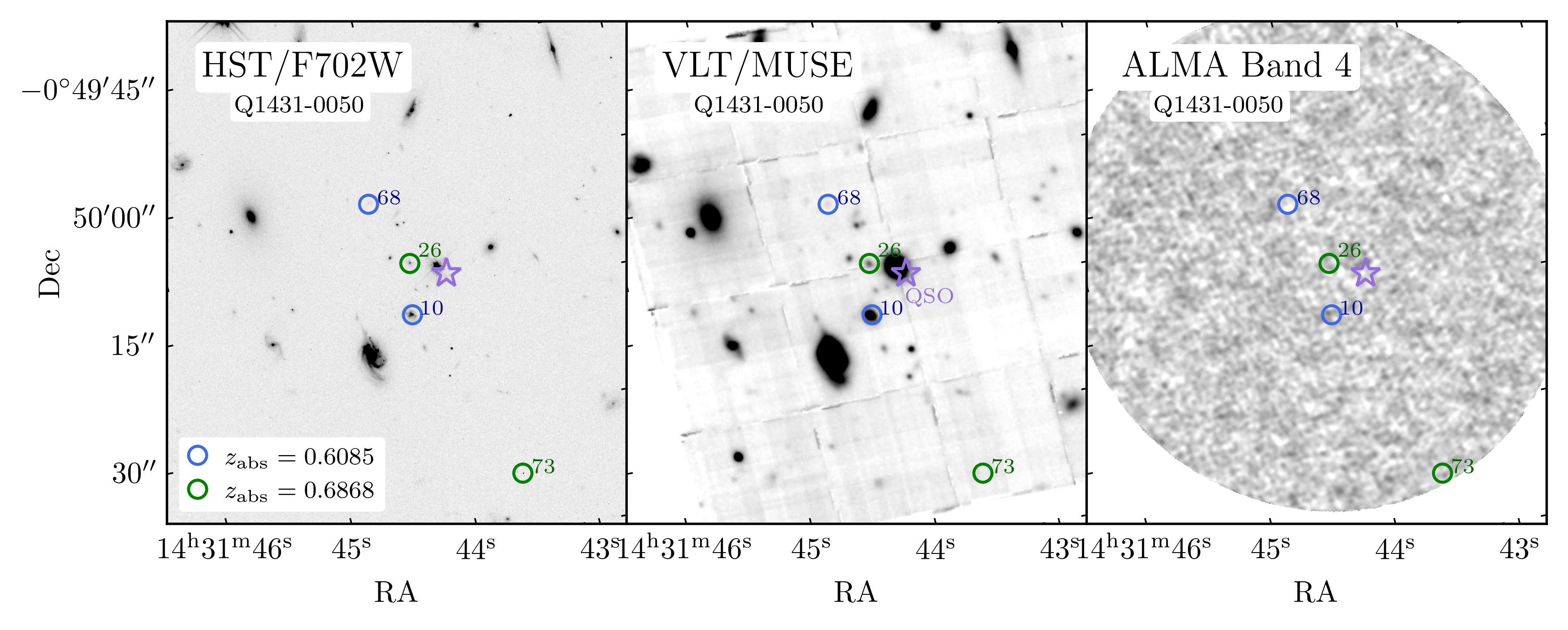}
\includegraphics[width=16.5cm, angle=0]{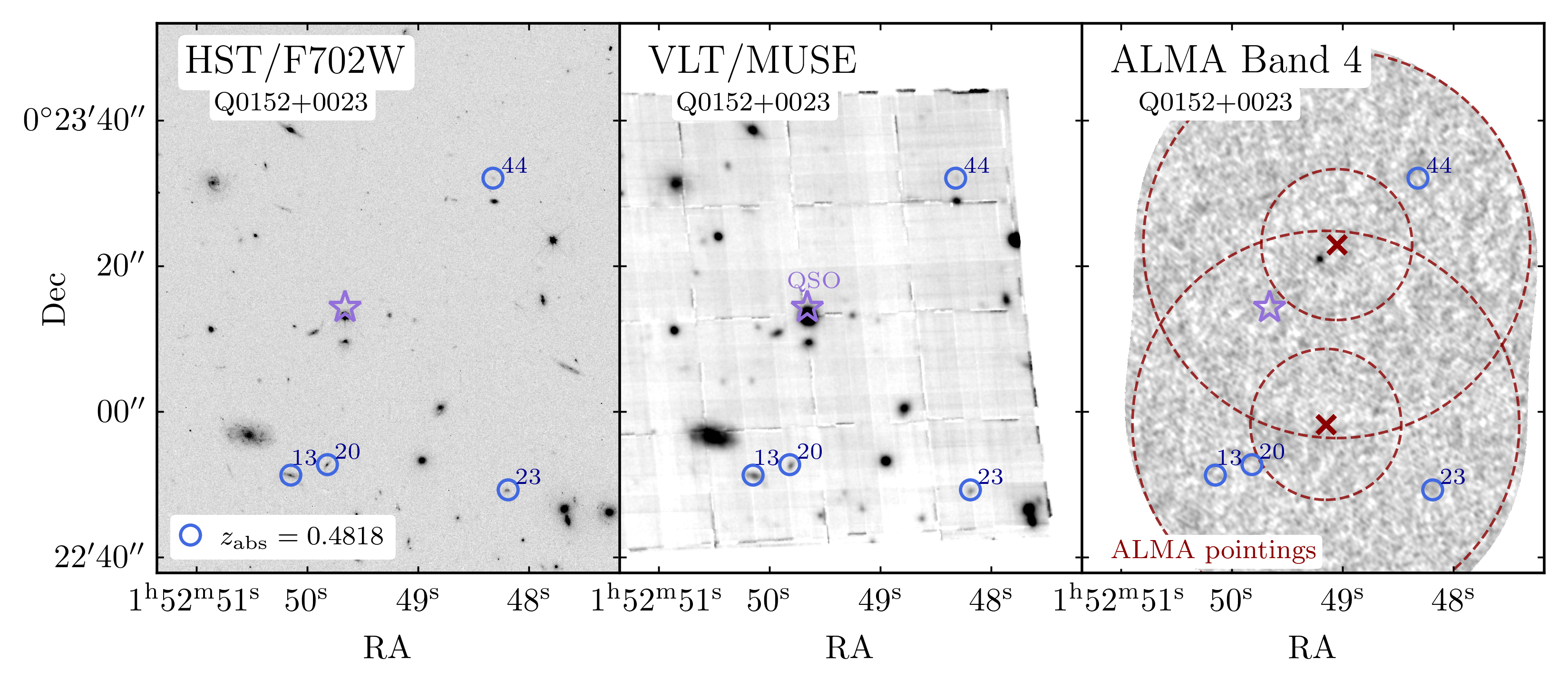}
  \caption {{ALMA Large Programme observing strategy.} Example observations of the targeted fields Q1431$-$0050 and Q0152$+$0023. {\it Left panel:} Broad-band HST/WFPC2 imaging displaying the exquisite spatial resolution afforded by space observations. The top row illustrates cases where they are distinct quasar absorbers at different redshifts with associated galaxies highlighted in blue and green. The purple star marks the position of the background quasar labeled "QSO" in the middle panel. {\it Middle panel:} VLT/MUSE white light of the same fields showing the position of the corresponding galaxies for which a spectrum is extracted. {\it Right panel:} ALMA observations of the same fields. The bottom row illustrates that in ALMA band 4, several overlapping pointings (shown as red circles) are used to cover all the galaxies associated with the absorbers and detected in HST and VLT/MUSE. The ALMA Large Programme observing strategy is thus optimised by targeting spectrally and spatially multiple VLT/MUSE-detected galaxies.  }
\label{fig:strategy}
\end{figure*}

The 61.5-hr ALMA Large Programme aims at characterising the molecular properties of 39 z$\sim$0.5 galaxies (see Table~\ref{tab:MAH_sample}) associated with absorbers with known \NHI. This efficient program simultaneously covers several absorbers at different frequencies and/or associated galaxies spatially as well as the background quasar. The strategy of including the quasar in the field-of-view improves the sensitivity of the observations by enabling self-calibration hence maximising the detection sensitivity. In addition, Figure~\ref{fig:strategy} provides examples where associated galaxies at 2 distinct redshifts (top row) were targeted and with multiple ALMA positions including several objects per pointings (bottom row). Table~\ref{tab:JoO_ALMA} summarises the number of galaxies and the number of pointings for each of the quasar absorbers. 

\begin{figure*}[h]
\includegraphics[width=21.cm, angle=0]{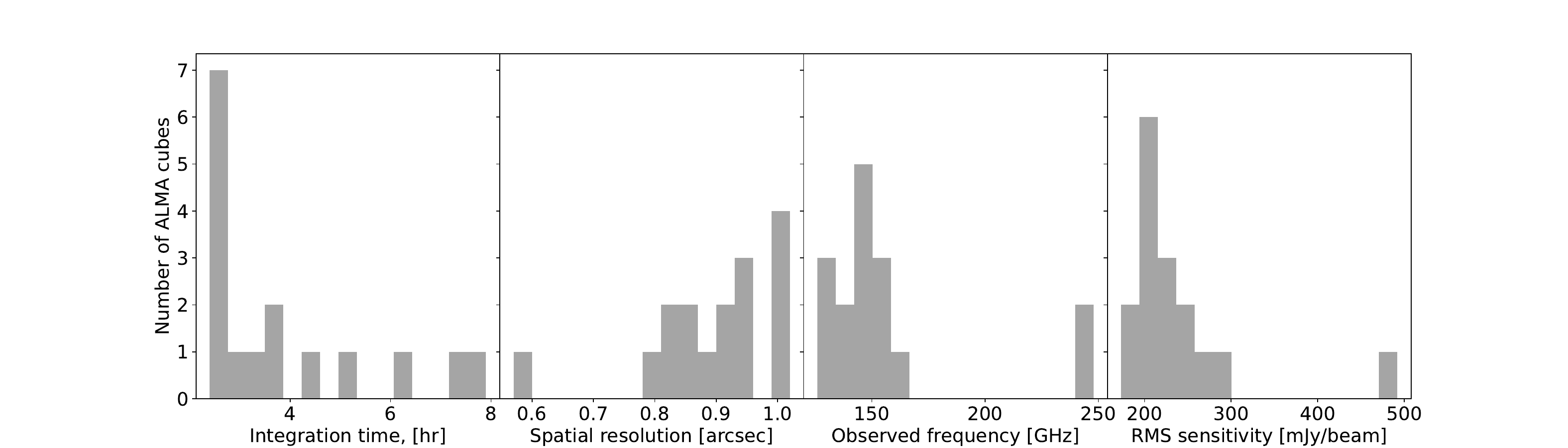}
  \caption {{ALMA Large Programme survey design.} The histograms show the distribution of exposure time (first panel), angular resolution (second panel), frequency (third panel), and noise in the continuum (fourth panel). These values are also tabulated in Table~\ref{tab:JoO_ALMA}. The overall survey design is rather homogeneous across quasar fields.  }
\label{fig:ALMAobs}
\end{figure*}

For the choice of ALMA bands, we devised an observing strategy based on minimising the number of set-ups while maximising the number of targets, and optimising atmospheric conditions (hence schedulability) and detection sensitivity. We targeted properly redshifted frequencies ($\sim$150GHz) to probe in band 4 CO(2$-$1) or in one occasion CO(3$-$2) ($z_{\rm abs}$=1.1532 towards Q0454$+$039). For two targets at higher redshifts $z_{\rm abs}$=0.8596 and 0.8999 (towards Q0454$+$039 and J1211$+$1030, respectively), we have observed CO(4$-$3) in band 6. Table~\ref{tab:JoO_ALMA} provides a summary of the bands and specific frequencies used for each quasar fields, which are also plotted in the third panel of Figure~\ref{fig:ALMAobs}.
The band 4 primary beam (FWHM $\sim$44 arcsec; $\sim$250kpc at the median sample redshift) corresponds to 75\% of MUSE field-of-view. 

All fields were set to be observable with a minimum two distinct ALMA configurations thus maximising the schedulability of the programme. The run made use of the configurations C-3, C-4 and some C-2. This was achieved by choosing a range of angular resolutions which have proven to enable kinematical studies of such targets \citep{Peroux19, Szakacs21, Barfety26}. We set angular resolution ranging 0.57-1.02 arcsec to optimise the forward-modelling of the velocity field and so as to enable comparison with the ionised gas probed in emission with MUSE and the neutral circumgalactic gas probed in absorption with VLT and Keck spectroscopy \citep{Rahmani18a}. Based on early ALMA observations, the sizes of targets were known to be significantly less than 5 arcsec so that there were no issues with missing emission which is more extended than the chosen largest recoverable scale (LAS) of 7 arcsec. The final list of angular resolutions achieved for each of the ALMA Large Programme fields can be found in Table~\ref{tab:JoO_ALMA} and its distribution is plotted on the second panel of Figure~\ref{fig:ALMAobs}.

The observations were designed to reach a molecular mass of log M$_{\rm H2}>$9 M$_\odot$ assuming a Milky Way Spectral Line Energy Distribution (SLED) to convert to CO(1$-$0) flux \citep{Carilli13}, a galactic conversion factor to estimate molecular mass of $\alpha_{CO}$=4.6 M$_\odot$ (K km/s pc$^2$)$^{-1}$ and the z=0.5 luminosity distance of dL=2129 Mpc using a (0.3,0.7,0.7) cosmology. The resulting exposure times ranged from 2.4 to 7.9 hrs, with most of the targets being observed with shortest intergration times as illustrated in the left panel of Figure~\ref{fig:ALMAobs}. In some cases, the quasar is bright at mm frequencies ($>$4 mJy), yet too faint to affect the dynamical range of the data. The resulting RMS continuum sensitivity does not exceed 500 $\mu$mJy/beam in all cases as illustrated in the right panel of Figure~\ref{fig:ALMAobs}.

\begin{table*}[tbh]
\begin{center}
\caption{{ The 17 quasar absorbers targeted with the ALMA Large Programme.} }
\vspace{0.2cm}
\begin{tabular}{lccccccccccc}
\hline \hline
Quasar 		 &$z_{\rm quasar}$  &$z_{\rm abs}$ &$\log$ & t$_{exp}$ &Resolution 
&ALMA & CO &Frequency  &\# galaxies &\# ALMA &Sensitivity \\
&&& \nhi&&&&&&&&RMS\\
                 &                &    &[cm$^{-2}$] &[hr]  &["] &Band & line & [GHz] &at $z_{\rm abs}$  & pointings  &[$\mu$Jy/beam] \\
\hline                                                        
J0152$-$2001 &2.06  &0.7802    &19.10    &3.7 &0.99	&4  &2-1  &129.52   &1  &1 &227  \\
Q0152$+$0023 &0.59  &0.4818    &19.78    &5.3 &0.79	&4  &2-1  &155.77   &4  &2 &206  \\
Q0454$+$039  &1.34  &0.8596    &20.67    &3.2 &0.57	&6  &4-3  &247.87   &1  &1 &492  \\
...	     &...   &1.1532    &18.59	 &2.8 &0.85	&4  &3-2  &153.68   &2  &1 &198  \\
Q0454$-$220  &0.53  &0.4744    &19.45    &2.4 &0.81	&4  &2-1  &156.83   &1  &1 &243  \\
...	     &...   &0.4833    &18.65	 &... &...	&4  &2-1  &155.77   &1  &...&...  \\
Q1110$+$0048 &0.76  &0.5604    &20.20    &2.4 &0.85	&4  &2-1  &147.78   &3  &1 &173   \\
J1211$+$1030 &2.19  &0.6296    &20.30    &2.7 &1.02	&4  &2-1  &141.43   &2  &1 &215   \\
...	     &...   &0.8999    &$<$18.50 &6.3 &0.87	&6  &4-3  &242.65   &3  &3 &292  \\
Q1229$-$021  &1.05  &0.8311    &18.84    &4.4 &1.01	&4  &2-1  &125.98   &4  &1 &219  \\
Q1342$-$0035 &0.79  &0.5380    &19.78    &2.4 &0.84	&4  &2-1  &149.89   &2  &1 &215   \\
Q1345$-$0023 &1.10  &0.6057    &18.85    &2.4 &0.95	&4  &2-1  &143.58   &2  &1 &268   \\
Q1431$-$0050 &1.18  &0.6085    &19.18    &2.4 &0.92	&4  &2-1  &144.09   &2  &1 &228   \\
...	     &...   &0.6868    &18.40	 &2.4 &0.94	&4  &2-1  &136.41   &2  &1 &245   \\
Q1515$+$0410 &1.27  &0.5592    &20.20    &7.3 &0.94	&4  &2-1  &147.86   &4  &2 &182   \\
Q1554$-$203  &1.95  &0.7869    &$<$19.0  &3.5 &1.00	&4  &2-1  &129.01   &1  &1 &199   \\
J2131$-$1207 &0.50  &0.4298    &19.50    &7.9 &0.92	&4  &2-1  &161.24   &4  &2 &206   \\      
\hline \hline
\label{tab:JoO_ALMA}
\end{tabular}			       			 	 
\begin{minipage}{180mm}
Note: The table lists the redshifts and neutral gas column densities, \nhi, as estimated from HST UV-spectra as well as the number of galaxies with redshift matching $z_{\rm abs}$, totalling to 39 objects. The penultimate column provides the number of ALMA pointings required to efficiently cover all these targets. 
The ALMA primay beam in band 4 is 44 arcsec while it is 27 arcsec in band 6.
\end{minipage}
\end{center}			       			 	 
\end{table*}

\subsection{Observations} 
\label{sec:obs}

The MUSE-ALMA Haloes ALMA Large Programme was observed as part of Cycle 10 under ID: 2023.1.00127.L. The observations took place between late December 2023 and May 2024. The observatory was extremely efficient in performing the observations so that the time between the ALMA panel review results are known to the proposers (Summer 2023) and the final observations are taken (Spring 2024) was remarkably short. 

The observations were performed with one baseband in the high resolution Frequency Division Mode (FDM), dual polarisation, but binned by a factor 4 to achieve a spectral resolution of 15 km/s over a total bandwidth of 3600 km/s to resolve the emission lines. The remaining three basebands are used in Time Division Mode (TDM) in order to achieve good continuum sensitivity.

A journal of observations summarising the properties of the ALMA Large Programme run is presented in Table~\ref{tab:JoO_ALMA}. The table includes exposure times, spatial resolution, targeted frequency coverage and sensitivity achieved for each of the observations. 

\subsection{Data reduction} \label{sec:redu}

The data reduction and imaging largely followed the ALMA observatory guidelines. Specifically, we extract the calibrated measurement sets ({\it ms}) produced with the Common Astronomy Software Applications (\textsc{CASA}, version 6.5.6) by the observatory. In addition, we apply self-calibration when possible and produce the images and datacubes with different weighting schemes and channel widths.

In six of the observed fields, the background quasars are bright enough at mm frequencies ($\geq$ 4 mJy) to perform self-calibration. Self-calibration corrects for short-term phase and amplitude variations, resulting in a marked improvement in the sensitivity hence dynamical range of the final images. We first clean the continuum image of the quasar field interactively to generate the clean model for self-calibration. During the first round of interactive cleaning, we clean the quasar based on the point source assumption. Then, we derive a phase-only calibration with various time intervals ranging from 30 seconds to 4 minutes, depending on the brightness of the quasar. We select the time intervals based on the improvements in the final imaging products. We also apply amplitude self-calibration if the quasar is brighter than 20 mJy and accepts the signals down to 5$\sigma$ when constructing the clean models. During self-calibration, the calibration tables are generated with the task \textsc{gaincal} and apply to the pipeline-calibrated measurements using \textsc{applycal}. After self-calibration, the RMS sensitivity is improved by factors ranging from 1.1 to 7.6, compared to the ones without self-calibration. Besides, our manual self-calibration also improves the sensitivity by a factor of about 1.2 compared to the automated observatory-based self-calibration pipeline products. In one of the fields, Q2131$-$1207, the quasar is bright leaving some artifacts across the field-of-view. We use the \textsc{CASA} task \textsc{uvcontsub} to remove the continuum contamination from the bright quasar while excluding the line-emitting channels of our primary targets. 

We then image the (self-)calibrated visibility into continuum images and 3D images (datacubes) with the CASA task \textsc{tclean}. For four of the quasar fields, several pointings are used to maximise the spatial coverage of the targets with known sky positions. The images of the mosaic fields are created by setting the gridder to "mosaic". The images are created with two different weighting schemes. The "natural" weighting is used to maximize the detection sensitivity, and the "Briggs" weighting (robust=0.5) is used to achieve an optimal balance between sensitivity and spatial resolution. We chose two distinct velocity resolutions. The 15 km/s resolution corresponds to the intrinsic spectral resolution. We further apply a spectral binning of four channels to improve the signal-to-noise ratio, resulting in a spectral resolution of 60 km/s. As a result, four different types of images and datacubes are produced, for each of the two weighting schemes and each two of the spectral resolution choices. The natural weighting with the 60 km/s products is better suited for searches for line emissions, while the Briggs weighting with different velocity resolutions is better suited for detailed kinematic analysis of the targets with strong line emissions.

The final RMS noise level of the cubes ranges from 180 to 500 mJy/beam and is typically $<$ 250 mJy/beam. The resulting angular resolution varies from 0.57 to 1.02 arcsec. The ALMA primary beam in band 4 is 44 arcsec, while it is 27 arcsec in band 6. The sensitivity and angular resolution of each field are listed in Table~\ref{tab:JoO_ALMA}.

\begin{figure*}[h]
\label{fig:detection}
\includegraphics[width=18.5cm, angle=0]{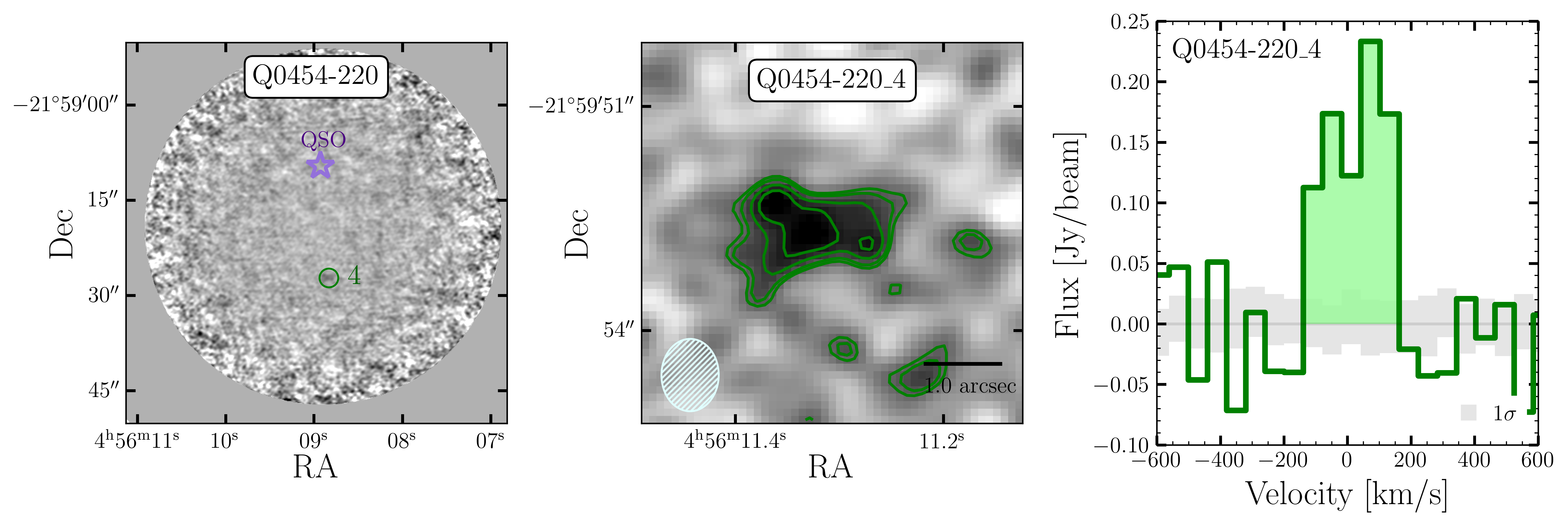}
  \caption{{ALMA Large Programme detection example.} {\it Left panel:} Full ALMA pointing for the Q0454$-$220 quasar field. The purple star marks the position of the background quasar labeled "QSO". {\it Middle panel:} Zoom-in on the CO(2--1) detection corresponding to the $z_{\rm abs}$=0.4833 absorber. The synthetised beam is indicated by the hashed area in the bottom left corner. {\it Right panel:} Corresponding ALMA spectrum displaying the flux emission at the expected frequency at 155.77 GHz. The detections of the full sample are reported in \cite{Bollo26}.}
\end{figure*}

The search for emission lines was performed on the datacubes without primary beam correction. This provides a homogeneous apparent noise level across the field-of-view. After we located the sources, their spectra were extracted from the primary beam-corrected datacubes generated by \textsc{impbcor}, which properly accounts for the increased noise level near the edge of the fields. The spectra were extracted within \textsc{CASA} using different aperture sizes and position angles to optimize the final SNR \citep{Bollo26}. { Fig.~\ref{fig:detection} displays an example of such spectrum. The left panel shows the whole ALMA field, while the CO(2--1) spatial and spectral detection is reported in the middle and right panels, respectively. }

\section{Release of data products} 
\label{sec:products}

This paper accompanies the data release of the MUSE-ALMA Haloes ALMA Large Programme datasets. Because of the legacy value of the programme, we aim at enabling science beyond the primary goals of the initial survey. In particular, the sample additionally provides deep high spatial resolution observations of mm-bright background quasars \citep{Girdhar22, Girdhar24, Molyneux24}. We therefore make the fully processed data available to the whole community roughly two years after the completion of the observing programme. The processed data include value-added data products generated by the team. The data products provided are:

\begin{itemize}

\item raw data files. 

\item one README file for each set of enhanced data uploaded (corresponding to a Group or Member OUS) in ascii (.txt) format following the observatory-provided template.

\item a detailed Data Description Document accompanying the Group (or Member) OUS enhanced data package, which provides the user with a complete list of all the uploaded data products and a description of their content. The document includes the project title and project code and provides detailed descriptions of data files (and their provenance) contained in the delivery. This file provides sufficient detail for a prospective user to reproduce the data reduction and understand the multiple data sets within the products. 

\item the full combined cubes including the .pbcor. FITS files according to ALMA standards, the corresponding 
.mask., and .pb. FITS files. The cubes are created with both natural weighting to maximize the signal-to-noise ratio and Briggs weighting with robust parameter set to 0.5. Both the spectral resolution of 15 km/s and the four-channel binned (corresponding to a spectral resolution of about 60 km/s) cubes which improved signal-to-noise are available. Cubes are supplied in FITS format.

\end{itemize}

While the raw data are available through the ALMA Science Archive\footnote{\url{https://almascience.eso.org/aq/}}, the remaining of the products are made available through a dedicated ALMA page\footnote{\url{www.eso.org/~cperoux/MUSE_ALMA_Haloes.html}}. These enhanced data products are “value added” and supplement the standard ALMA products that are generated by the ALMA observatory and already stored in the ALMA Science Archive.

In addition, the present paper summarises in Section~\ref{sec:MAH} the "primary targets" absorption characteristics (redshift, \hi\ column density and neutral gas metallicity) and emission properties (impact parameter, SFR, stellar mass and ionised gas metallicity). These values are listed in Table~\ref{tab:abs} in the Appendix.

\section{Science goals of the ALMA large programme} 
\label{sec:science}

The MUSE-ALMA Haloes Large Programme aims at characterising the molecular gas phase of galaxies with robustly determined neutral atomic \& ionised circumgalactic gas and stellar components. The overreaching goal is to constrain the full baryon cycle of a sample of z$\sim$0.5 galaxies. The study tackles the following key questions:

\begin{itemize}
\item What role does molecular gas play in \hi-rich galaxies? (Section \ref{sec:mol_gas})
\item How do molecular gas kinematics relate to gas flows? (Section \ref{sec:kine})
\item What is the condensed baryon census of the circumgalactic medium?  (Section \ref{sec:baryon_census})
\end{itemize}

\subsection{Quantify the molecular gas content of H\,I-rich galaxies} 
\label{sec:mol_gas}

One of the key findings of past observations with integral field spectroscopy is that strong N(H\,I) absorbers are associated with multiple galaxies detected with both VLT/MUSE and ALMA \citep{hamanowicz2020, Dutta20, Weng23a}. This important conclusion is also supported by
hydridynamical cosmological RAMSES zoom-in and TNG simulations demonstrating that the physical properties of these objects are qualitatively similar to a small group environment, possibly part of a cosmic filament \citep{Peroux19, augustin2018, Weng24}. This challenges the historical view of strong-N(H\,I) quasar absorbers being associated with a single bright galaxy. Instead, several lines of evidence suggest a situation where the H\,I gas probed in absorption is related to complex galaxy structures \citep{Klitsch18, hamanowicz2020}. In addition to MUSE detections, ALMA observations have provided new information on the molecular gas content of these systems over the last decade. Building on these successes, the MUSE-ALMA Haloes survey ALMA Large Programme characterises the molecular phase of the neutral gas of a sample of galaxies associated with strong H\,I-absorbers. This dataset quantifies:

i) {\it CO in \hi-selected galaxies} - we efficiently target 39 galaxies with known sky positions and measured redshifts. What makes this survey unique with respect to other projects, are accurate estimates of the SFR, metallicity and stellar mass of all targets. Early studies \citep{Peroux19, Szakacs21} have demonstrated that the powerful combination of ALMA and VLT/MUSE observations is pivotal to securely assess the redshift of single-line mm detections. The ALMA Large Programme enables stacking analysis of emission lines to reach even lower flux limits { of typically 250 $\mu$Jy km/s}. These results can be contrasted with the latest hydrodynamical cosmological simulations including GRIFFIN \citep{Lahen20, Szakacs22}, COLDSim \citep{Maio22, Casavecchia24, Casavecchia25, Maio26} or COLIBRE \citep{Schaye25, Chaikin25, Trayford26} to constrain which physical conditions (in terms of temperature and density) of the gas CO best traces.

ii) {\it Detection rate} - some studies have focused their search on metal-rich absorbers \citep{Neeleman16, Kanekar18, Kanekar20}, whilst the sample targeted here has no prior on metallicity. The ALMA Large Programme allows us to assess the frequency of CO emitters down to log M$_{\rm H2}>$9 M$_\odot$ among H\,I-selected galaxies using an unbiased sample, selected irrespective of their metallicity. It additionally enables the assessment of the incidence of "CO-dark" molecular gas, defined as regions where H$_2$ exists without detectable CO emission due to
low metallicity, reduced dust shielding, or diffuse, extended gas distributions \citep{Bollo26}. In case of non-detections, the ALMA Large Programme is able to establish whether the ionised gas properties of CO detections and non-detections are different.

\subsection{Characterise multiphase gas flows} 
\label{sec:kine}

Galaxy-scale processes through which galaxies accrete and convert their gas into stars need to be studied in
selected objects in order to obtain a complete picture of galaxy formation. In particular importance are the interactions
between gas accretion (imposed by large-scale dark matter structures from IGM reservoirs) and feedback
(launched by star-forming regions in galaxies and active supermassive black holes). It is thus key to study gas flows into and out of galaxies simultaneously as their
mutual interactions affect each other. In particular, characterising the physical processes of the inflows of gas accreting into galaxies and the outflows carrying away the energy and metals generated inside galaxies. In this context, the MUSE-ALMA Haloes survey ALMA Large Programme establishes the:

i) {\it Kinematics of molecular gas} - forward modelling of the 3-dimensional morpho-kinematics allows us to measure the gas circular velocity at the position of the background quasar sighline. Software tools for such studies include GalPak$^{\rm 3D}$ \citep{Bouche15, Bouche22}, LINEFIT \citep{Davies11} and DysmalPy \citep{Davies04a,Davies04b, Cresci09, Wuyts16, Lang17, Price21, LeeLilian25}. Dynamical mass estimates reveal the physics behind the build-up of intermediate redshift galaxies. The ALMA Large Programme indeed gathers new information on the orientation, geometry and kinematics of these galaxies. These observations, in concert with the absorption profile along the sightline to the background quasar, enable identifying signatures of gas flows departing from disk rotation. 

ii) {\it Ionised and molecular gas kinematics} - VLT/MUSE and ALMA observations combined offer an unprecedented view on the kinematics of the ionised and molecular gas at comparable spatial and spectral resolutions. The data sets also provide a still rare opportunity to characterize the galactic cold molecular and warm ionised gas kinematics {\em in the same} galaxies, to address in particular the issue of whether the associated gas components have similar local velocity dispersion with implications on the physical driver of gas turbulence \citep[feedback- and/or gravity-driven; e.g.][]{Krumholz18, Ginzburg22}. Specifically, by comparing the outflow velocities to SFRs, dynamical, and stellar masses, the ALMA Large Programme investigates if inflow and outflow velocities differ in molecular and ionised gas \citep{Carniani16, HerreraCamus25}.

\subsection{A census of the circumgalactic medium condensed baryons}
\label{sec:baryon_census}

\cite{peroux2020} refer to the condensed matter as to the sum of stars, dust, neutral and molecular gas components. The cosmic evolution of its density is well measured by current observations. \cite{peroux2020} find that the observed z $<$ 2 decline of the star-formation history is caused by the lack of molecular gas supply related to a drop in net gas accretion rate. This decline is itself driven by the decreased growth of the dark matter halos. These observational results are in stark agreement with the gas
regulator model \citep{Lilly13, peng2014}, which describes a continued cycling of baryons flowing in and out of galaxies regulating
galaxy evolution. Interestingly, the neutral gas density dominates over that of the molecular gas at all redshifts  on global scales \citep{Decarli16, Walter20, Hamanowicz23, Bollo25}. 

These results need to be reconciled with measurement of the \hi\ and H$_2$ gas content of galaxy on local scales. Indeed, galaxy halos appear to be hiding approximately 60\% of their baryons compared to expectations from the cosmological mass density \citep{mcgaugh2008, werk14}. With the wealth of multi-wavelength information already available in the MUSE-ALMA Haloes survey, we are in a unique position to perform a full census of the condensed baryons in these objects \citep{OBeirne26}. The ALMA Large Programme characterises the missing component: the molecular content of these galaxies. As such the ALMA Large Programme allows us to probe the:

i) {\it Molecular mass} - the observations enable estimates of the molecular masses from CO emission flux measurements, using updated Spectral Line Energy Distribution \citep{Klitsch22} and CO-to-H$_2$ conversion factors \citep{Genzel15}. 
Combined with known SFR estimates, these observations enable calculation of the rate of conversion of gas into stars, the so-called molecular depletion times following: $\tau_{\rm depl}$=M$_{\rm H2}$/SFR=1/SFE, where SFE is the star formation efficiency, a key link to the baryon cycle. Recent studies of outer disk rotation curves at high redshift indicate low galactic-scale dark matter mass fractions f$_{\rm DM}$ among massive star-forming disks, with higher fractions derived towards lower masses \citep{Genzel17, Ubler18, Genzel20, Price21, Szakacs21, Bouche22, NestorShachar23, Puglisi23}. The ALMA Large Programme addresses this puzzle and uniquely connects the molecular kinematics to the properties of the dark matter halos at z$\sim$0.5 and over a wide dynamic range in galaxy mass. These observational results can be contrasted with expectations of hydrodynamical cosmological simulations such as TNG50 \citep{Pillepich19}.

ii) {\it Baryonic and gas fraction} - most recent CO-emission surveys such as COLD GASS \citep{Saintonge17} or the 2400 objects reported by PHIBBS \citep{Tacconi18, Tacconi20} have informed us on the gas fraction of galaxies up to z$\sim$4. 
The ALMA Large Programme allows us to characterise the gas fraction of 
less massive galaxies
by reaching objects with stellar masses M$_*$$<$10$^{9}$ M$_{\odot}$.

\section{Conclusions}\label{sec:conclusions}

A complete understanding of the baryon cycle necessitates a thorough inventory of condensed matter, comprised of stars and cold gas, within both galaxies and their circumgalactic medium. Today, the neutral atomic gas \hi\ is best probed by the emerging technique of examining the gas in absorption against background sources whose lines of sight pass through an identified foreground galaxy's halo. Recent advancements in 3D Integral Field Spectroscopy, which generates data cubes with a spectrum for each image pixel, have significantly improved the detection of ionised gas emission in the circumgalactic medium of these foreground galaxies at both optical and mm wavelengths.

The MUSE-ALMA Haloes survey builds on these techniques to yield several key results regarding the connection between galaxies and their CGM. The survey has found a high detection rate of galaxies associated with H I absorbers, with 89\% of the targeted absorbers having at least one associated galaxy within an impact parameter of 250 kpc \citep{Peroux17, Peroux19, hamanowicz2020, Weng23a}. A significant finding is that most of these absorption systems are associated with pairs or groups of galaxies, indicating the importance of the galaxy environment in shaping the CGM \citep{hamanowicz2020}. Interestingly, the galaxies with the smallest impact parameters to the quasar sightline are not always the closest in velocity space to the absorbing gas \citep{hamanowicz2020}. There is an anti-correlation between the H I column density of the absorbers and the impact parameter (scaled by the virial radius) of the nearest galaxy, which is consistent with predictions from simulations \citep{Weng23a}. The metallicity of the associated galaxies is typically higher than the metallicity of the absorbers. The absorber metallicity is also found to decrease at larger impact parameters from the host galaxy \citep{Weng23a}. The study found marginal evidence for a bimodal distribution in the azimuthal angles of strong H I absorbers, which is similar to what has been observed for other absorption lines like Mg II and O VI \citep{Weng23b}.
Detailed gas kinematics studies have provided unique evidence of gas flows in the CGM of galaxies, both winds \cite{Rahmani18b} as well as accretion \citep{Rahmani18a, Szakacs21}. A case-by-case analysis of the absorbers indicates that about 10\% are consistent with gas accretion, while up to 30\% trace outflows. The remaining absorbers are associated with gas in the galaxy disk, the intragroup medium, or low-mass galaxies that are below the detection limit of the MUSE instrument \citep{Weng23b}. 

Using HST broad-band imaging observations, \cite{Karki23} report that SFR shows a strong correlation with stellar mass, indicating that most absorption-selected galaxies align with the star formation main sequence of the global galaxy population. In addition, they find that more massive absorber counterparts exhibited higher central concentration and larger sizes. Finally, they report that the SFR surface densities of the absorption-selected galaxies were higher than predicted by the Kennicutt–Schmidt relation for local galaxies, which points to a higher star formation efficiency in these galaxies \citep{Karki23}.By fitting the spectral energy distribution from HST data, the survey has determined the stellar masses of the galaxies associated with the absorbers, which span a wide range from 8.1 $< \log(M_\star /M_{\odot}) <$ 11.8 \citep{Augustin24}. Most of these galaxies are found to lie on the main sequence of star-forming galaxies \citep{Augustin24}.

Early ALMA observations have enabled the study of the molecular gas in these galaxies. First findings have found that \hi-selected galaxies tend to have high molecular gas masses given their low star formation rates, leading to high depletion times \citep{Klitsch18, Peroux19, Szakacs21}. In this paper, we present the Cycle 10 ALMA Large Programme campaign's which optimised setup enables us to study objects down to log M$_{\rm H2}>$9 M$_\odot$ at 5-$\sigma$ thus providing a large sample required to measure evolution and determine typical properties of these systems. The paper provides a description of the survey design, details on the observational set-up and an extensive report of data reduction processes. It also accompanies the data release of the ALMA Large Programme. Finally, the paper reviews the salient science goals that will be tackled with these extensive datasets. The project increases the number of galaxies with observational constraints of their multi-phase circumgalactic medium by more than a factor 2, thus producing a representative sample of \hi-rich galaxies. The overarching objective of the project is to characterise the multi-phase properties of a remarkable sample of galaxies with multi-wavelength observations with VLT/MUSE, HST UV spectroscopy and broad-band imaging to probe the physical processes by which gas is transformed into stars. By measuring the molecular mass, gas flows and gas fractions of this unique sample, the results will: 1) quantify the amount of molecular gas in \hi-rich galaxies, 2) characterise the role of molecular gas flows from kinematics analysis, and 3) perform a first census of the condensed baryons in the circumgalactic medium. Given the \hi-selection, these data will also prepare for future observations with SKA, including MeerKAT and ASKAP pathfinders.

\begin{acknowledgements}
     This research was supported by the International Space Science Institute (ISSI) in Bern, through ISSI International Team project \#564 (The Cosmic Baryon Cycle from Space). J.C., C.B., and N.M.F.S. acknowledge funding by the European Union (ERC Advanced Grant GALPHYS, 101055023). R.A. acknowledges funding from the European Research Council (ERC) under the European Union’s Horizon 2020 research and innovation program (grant agreement 101020943, SPECMAP-CGM). Views and opinions expressed are, however, those of the author(s) only and do not necessarily reflect those of the European Union or the European Research Council. Neither the European Union nor the granting authority can be held responsible for them. V.P.K. gratefully acknowledges support from NSF grants 2007538 and 2009811. This paper makes use of the following ALMA data: ADS/JAO.ALMA\#2023.1.00127.L ALMA is a partnership of ESO (representing its member states), NSF (USA) and NINS (Japan), together with NRC (Canada), MOST and ASIAA (Taiwan), and KASI (Republic of Korea), in cooperation with the Republic of Chile. The Joint ALMA Observatory is operated by ESO, AUI/NRAO and NAOJ. In addition, publications from NA authors must include the standard NRAO acknowledgement: The National Radio Astronomy Observatory is a facility of the National Science Foundation operated under cooperative agreement by Associated Universities, Inc.

\end{acknowledgements}

\section*{Data availability}

ALMA data directly related to this publication and its figures are publicly available. Additionally, the raw data can be downloaded from the public archives with the respective project codes.

%
  \bibliographystyle{aa} 
\bibliography{aanda}
%

\begin{appendix}

\section{MUSE-ALMA haloes survey publications to date}

{This appendix introduces Table~\ref{tab:publi} which lists the MUSE-ALMA Haloes survey publications to date.} 

\begin{tabular}{ll}
\hline\hline
Reference & Paper Title\\
\hline
\cite{Peroux17}	& Nature of the absorbing gas associated with a galaxy group at z$\sim$0.4\\
\cite{Rahmani18a}& Observational signatures of a warped disk associated with cold-flow accretion\\
\cite{Rahmani18b}& A Lyman limit system associated with galactic winds\\
\cite{Klitsch18}& A combined ALMA and MUSE survey for neutral, molecular, and ionized gas\\
\cite{Peroux19}	& Multiphase circumgalactic medium probed with MUSE and ALMA\\
\cite{hamanowicz2020} & MUSE-ALMA Haloes V: physical properties and environment of z$\leq$1.4 HI quasar absorbers\\
\cite{Szakacs21} & MUSE-ALMA Haloes VI: coupling atomic, ionized, and molecular gas kinematics of galaxies\\
\cite{Peroux22} & MUSE-ALMA Haloes VII: survey science goals \& design, data processing and final catalogues\\
\cite{Weng23a} & MUSE-ALMA Haloes VIII: Statistical study of circumgalactic medium gas\\
\cite{Karki23} & MUSE-ALMA Haloes IX: Morphologies and stellar properties of gas-rich galaxies\\
\cite{Augustin24} & MUSE-ALMA Haloes X: the stellar masses of gas-rich absorbing galaxies\\
\cite{Weng23b} & MUSE-ALMA Haloes XI: gas flows in the circumgalactic medium\\
\cite{Weng24} & The physical origins of gas in the CGM using observationally motivated TNG50 mocks\\
\cite{Bollo26} & MUSE-ALMA Haloes XII: Molecular gas in z$\sim$0.5 HI-selected galaxies \\
\cite{Karki26} & MUSE-ALMA Haloes XIII: Kinematics and Stellar Properties of Tidal Features \\
\cite{Peroux26} & MUSE-ALMA Haloes XIV: The ALMA Large Programme Data Release\\
\cite{OBeirne26} & MUSE-ALMA Haloes XV: A census of the condensed baryons in $z\sim0.5$ galaxies\\
\cite{Barfety26} & MUSE-ALMA Haloes XVI: Multiphase kinematics and rotational support of different gas phases\\
\cite{Halley26} & MUSE-ALMA Haloes XVII: Circumgalactic Medium metallicity\\
\hline\hline 				       		         	 
\label{tab:publi}                                         
\end{tabular}

\section{MUSE-ALMA Haloes survey associated galaxy properties}
This appendix presents the table summarising the physical properties of all 79 galaxies associated with the 32 MUSE-ALMA Haloes primary targeted quasar absorbers. 

\onecolumn
\noindent
\begin{longtable}{llccccccc}
\caption{{MUSE-ALMA Haloes Survey associated galaxy properties.}  }
\label{tab:abs} \\
\hline
Quasar/Gal &$z_{\rm abs}^{1}$ &\logNHI$^{1}$ &absorption &$z_{\rm gal}^{3}$ &impact & SFR$^{3}$ & stellar  &emission \\
&&&metallicity$^{2}$ &&parameter$^{3}$ &&mass$^{4}$ &metallicity$^{3}$\\
&&[atoms/cm$^2$] &[M/H] & &[kpc] &[M$_{\odot}$/yr] &[M$_{\odot}$] &$12 + \log$(O/H)\\
\hline
\endfirsthead
\caption{continued}\\
\hline
\endhead
\hline
\endfoot
Q0058p0019   	      	&0.6125 &20.08$\pm$0.15	&0.25$^{+0.19}_{-0.19}$		&... 	     	&...  	&...  	        	&...		      &... 	  	 \\	  		
Q0123m0058   	      	&0.8686 &$<$18.62	&$<$-1.0			&... 	     	&...  	&...  	        	&...		      &... 	  	 \\	  		
Q0123m0058   	      	&1.4094 &20.08$\pm$0.09	&-0.6$^{+0.004}_{-0.01}$	&... 	     	&...  	&...  	        	&...		      &... 	  	 \\	  		
Q0138m0005$\_$14$^b$	&0.7821	&19.81$\pm$0.08	&0.28$^{+0.15}_{-0.15}$		&0.78212  	&81.5	&6.9  $\pm$2.44	   	& 9.8$_{-0.2}^{+0.1}$ &8.72$\pm$0.07	 \\
Q0152m2001$\_$4		&0.3830	&$<$18.8	&$<$-1.5			&0.38138  	&169.6	&2.05 $\pm$0.66	   	&11.5$_{-0.1}^{+0.1}$ &8.73$\pm$0.1	 \\
Q0152m2001$\_$5$^c$	&0.3830	&$<$18.8	&$<$-1.5			&0.38257  	&59.9	&0.1  $\pm$0.38	   	&11.3$_{-0.1}^{+0.1}$ &8.65$\pm$0.08	 \\
Q0152m2001$\_$7		&0.3830	&$<$18.8	&$<$-1.5			&0.38135  	&146.5	&0.16 $\pm$0.42	   	&11.0$_{-0.1}^{+0.1}$ &8.49$\pm$0.15	 \\
Q0152m2001$\_$13$^a$	&0.3830	&$<$18.8	&$<$-1.5			&0.38145  	&84.2	&$<$0.028		&11.8$_{-0.4}^{+0.2}$ &... 	  	 \\
Q0152m2001$\_$14	&0.3830	&$<$18.8	&$<$-1.5			&0.38222  	&204.4	&0.73 $\pm$0.37	   	&10.5$_{-0.2}^{+0.2}$ &8.58$\pm$0.08	 \\
Q0152m2001$\_$62	&0.3830	&$<$18.8	&$<$-1.5			&0.38498  	&144.2	&$<$0.028		&10.5$_{-0.1}^{+0.1}$ &... 	  	 \\
Q0152m2001$\_$12$^a$	&0.7802	&18.87$\pm$0.12	&-1.9$^{+0.1}_{-0.01}$		&0.78018  	&55.0	&0.18 $\pm$0.4	   	&                 ... &... 	  	 \\
Q0152p0023$\_$7		&0.4818	&19.78$\pm$0.08	&-0.1$^{+0.011}_{-0.019}$	&0.48261  	&154.3	&2.6  $\pm$0.96	   	&10.1$_{-0.1}^{+0.2}$ &8.68$\pm$0.1	 \\
Q0152p0023$\_$13	&0.4818	&19.78$\pm$0.08	&-0.1$^{+0.011}_{-0.019}$	&0.48206  	&140.0	&2.7  $\pm$0.86	   	& 9.0$_{-0.1}^{+0.2}$ &8.43$\pm$0.06	 \\
Q0152p0023$\_$20$^a$	&0.4818	&19.78$\pm$0.08	&-0.1$^{+0.011}_{-0.019}$	&0.48143  	&122.9	&0.83 $\pm$0.35	   	& 8.1$_{-0.1}^{+0.1}$ &8.59$\pm$0.12	 \\
Q0152p0023$\_$23$^a$	&0.4818	&19.78$\pm$0.08	&-0.1$^{+0.011}_{-0.019}$	&0.48068  	&187.8	&$<$0.243		& 8.4$_{-0.1}^{+0.1}$ &... 	  	 \\
Q0152p0023$\_$44$^a$	&0.4818	&19.78$\pm$0.08	&-0.1$^{+0.011}_{-0.019}$	&0.4832	  	&170.7	&$<$0.243		& 8.0$_{-0.6}^{+0.6}$ &... 	  	 \\
Q0420m0127$\_$8$^d$	&0.6331	&18.54$\pm$0.09	&-1.6$^{+0.1}_{-0.01}$		&0.63318  	&133.9	&1.03 $\pm$0.36	   	&10.3$_{-0.4}^{+0.2}$ &8.59$\pm$0.12	 \\
Q0420m0127$\_$12$^d$	&0.6331	&18.54$\pm$0.09	&-1.6$^{+0.1}_{-0.01}$		&0.63382  	&144.8	&1.24 $\pm$0.42	   	&10.8$_{-0.1}^{+0.1}$ &8.54$\pm$0.14	 \\
Q0420m0127$\_$13$^d$	&0.6331	&18.54$\pm$0.09	&-1.6$^{+0.1}_{-0.01}$		&0.63326  	&101.4	&0.83 $\pm$0.3		& 9.7$_{-0.9}^{+0.9}$ &8.44$\pm$0.18	 \\
Q0420m0127$\_$30	&0.6331	&18.54$\pm$0.09	&-1.6$^{+0.1}_{-0.01}$		&0.63197  	&215.8	&2.68 $\pm$0.87	   	& 9.1$_{-0.9}^{+0.9}$ &... 	  	 \\
Q0454m220$\_$69$^a$	&0.4744	&19.45$\pm$0.03	&-0.48$^{+0.1}_{-0.1}$		&0.4745	  	&8.4	&4.26 $\pm$1.32	   	&10.8$_{-0.1}^{+0.1}$ &8.66$\pm$0.05	 \\
Q0454m220$\_$4$^a$	&0.4833	&18.65$\pm$0.02	&-0.7$^{+0.1}_{-0.7}$		&0.48377  	&107.6	&0.68 $\pm$0.27	   	&10.6$_{-0.1}^{+0.1}$ &8.33$\pm$0.12	 \\
Q0454p039$\_$57$^a$	&0.8596	&20.69$\pm$0.06	&-1.57$^{+0.17}_{-0.17}$	&0.85877  	&18.3	&7.64 $\pm$2.35	   	&                 ... &... 	  	 \\
Q0454p039$\_$15$^a$	&1.1532	&18.59$\pm$0.02	&-0.19$^{+0.22}_{-0.22}$	&1.15497  	&58.0	&0.65 $\pm$0.22	   	&                 ... &... 	  	 \\
Q0454p039$\_$65$^a$	&1.1532	&18.59$\pm$0.02	&-0.19$^{+0.22}_{-0.22}$	&1.1547	  	&127.1	&0.71 $\pm$0.25	   	& 9.9$_{-0.8}^{+0.8}$ &... 	  	 \\
Q1110p0048$\_$6$^a$	&0.5604	&20.20$\pm$0.10	&...				&0.56006  	&59.6	&$<$0.122		& 9.1$_{-0.1}^{+0.1}$ &8.26$\pm$0.09	 \\
Q1110p0048$\_$15$^a$	&0.5604	&20.20$\pm$0.10	&...				&0.56035  	&129.7	&$<$0.122		& 8.8$_{-0.6}^{+0.6}$ &8.46$\pm$0.1	 \\
Q1110p0048$\_$44$^a$	&0.5604	&20.20$\pm$0.10	&...				&0.56118  	&5.7	&$<$0.122		& 9.2$_{-0.3}^{+0.3}$ &8.49$\pm$0.05	 \\
Q1130m1449$\_$18	&0.1906	&$<$19.1	&$<$-2.3			&0.19048  	&17.9	&0.33 $\pm$0.33	   	&10.9$_{-0.1}^{+0.1}$ &8.52$\pm$0.06	 \\
Q1130m1449$\_$4$^c$	&0.3127	&21.71$\pm$0.07	&-1.23$^{+0.16}_{-0.16}$	&0.31312  	&45.0	&0.56 $\pm$0.33	   	&                 ... &8.37$\pm$0.06	 \\
Q1130m1449$\_$5$^c$	&0.3127	&21.71$\pm$0.07	&-1.23$^{+0.16}_{-0.16}$	&0.31276  	&81.6	&0.68 $\pm$0.31	   	&10.9$_{-0.1}^{+0.1}$ &8.27$\pm$0.06	 \\
Q1130m1449$\_$8$^c$	&0.3127	&21.71$\pm$0.07	&-1.23$^{+0.16}_{-0.16}$	&0.31138  	&98.6	&0.04 $\pm$0.34	   	&10.5$_{-0.1}^{+0.1}$ &8.57$\pm$0.05	 \\
Q1130m1449$\_$13$^e$	&0.3127	&21.71$\pm$0.07	&-1.23$^{+0.16}_{-0.16}$	&0.31184  	&120.0	&$<$0.018		& 9.9$_{-0.2}^{+0.2}$ &... 	  	 \\
Q1130m1449$\_$16$^e$	&0.3127	&21.71$\pm$0.07	&-1.23$^{+0.16}_{-0.16}$	&0.3135	  	&37.3	&0.02 $\pm$0.59	   	& 9.3$_{-0.2}^{+0.3}$ &... 	  	 \\
Q1130m1449$\_$17$^e$	&0.3127	&21.71$\pm$0.07	&-1.23$^{+0.16}_{-0.16}$	&0.31197  	&17.6	&0.1  $\pm$0.33	   	& 9.2$_{-0.4}^{+0.4}$ &8.47$\pm$0.07	 \\
Q1130m1449$\_$24$^e$	&0.3127	&21.71$\pm$0.07	&-1.23$^{+0.16}_{-0.16}$	&0.31249  	&17.9	&2.08 $\pm$0.42	   	& 8.9$_{-0.2}^{+0.2}$ &8.37$\pm$0.11	 \\
Q1130m1449$\_$31$^e$	&0.3127	&21.71$\pm$0.07	&-1.23$^{+0.16}_{-0.16}$	&0.31267  	&73.4	&0.1  $\pm$0.34	   	& 8.6$_{-0.3}^{+0.2}$ &8.5 $\pm$0.07	 \\
Q1130m1449$\_$36$^e$	&0.3127	&21.71$\pm$0.07	&-1.23$^{+0.16}_{-0.16}$	&0.31322  	&23.2	&0.45 $\pm$0.6		& 8.8$_{-0.4}^{+0.4}$ &... 	  	 \\
Q1130m1449$\_$43$^e$	&0.3127	&21.71$\pm$0.07	&-1.23$^{+0.16}_{-0.16}$	&0.3124	  	&26.8	&$<$0.018		&                 ... &... 	  	 \\
Q1130m1449$\_$50	&0.3127	&21.71$\pm$0.07	&-1.23$^{+0.16}_{-0.16}$	&0.31437  	&99.3	&0.01 $\pm$0.49	   	&                 ... &8.45$\pm$0.15	 \\
Q1130m1449$\_$56$^e$	&0.3127	&21.71$\pm$0.07	&-1.23$^{+0.16}_{-0.16}$	&0.31215  	&129.6	&2.77 $\pm$0.37	   	&                 ... &8.7 $\pm$0.05	 \\
Q1130m1449$\_$68$^e$	&0.3127	&21.71$\pm$0.07	&-1.23$^{+0.16}_{-0.16}$	&0.31299  	&74.7	&0.02 $\pm$0.61	   	& 8.2$_{-0.3}^{+0.3}$ &8.44$\pm$0.17	 \\
Q1130m1449$\_$76$^e$	&0.3127	&21.71$\pm$0.07	&-1.23$^{+0.16}_{-0.16}$	&0.31295  	&9.9	&0.01 $\pm$0.44	   	& 7.9$_{-0.4}^{+0.4}$ &8.43$\pm$0.15	 \\
Q1130m1449$\_$6		&0.3283	&$<$18.9	&$<$-1.1			&0.3284	  	&91.1	&1.39 $\pm$0.37	   	& 7.8$_{-0.4}^{+0.4}$ &8.66$\pm$0.06	 \\
Q1130m1449$\_$9		&0.3283	&$<$18.9	&$<$-1.1			&0.32813  	&75.5	&1.02 $\pm$0.36	   	&                 ... &8.69$\pm$0.05	 \\
Q1211p1030$\_$7$^c$	&0.3929	&19.46$\pm$0.08	&-2.55$^{+0.94}_{-0.94}$	&0.39277  	&38.2	&3.0  $\pm$1.18	   	& 8.8$_{-0.1}^{+0.1}$ &... 	  	 \\
Q1211p1030$\_$9		&0.3929	&19.46$\pm$0.08	&-2.55$^{+0.94}_{-0.94}$	&0.39288  	&39.0	&3.87 $\pm$1.26	   	& 9.7$_{-0.4}^{+0.4}$ &... 	  	 \\
Q1211p1030$\_$17	&0.3929	&19.46$\pm$0.08	&-2.55$^{+0.94}_{-0.94}$	&0.39289  	&36.8	&$<$0.035		& 8.7$_{-0.1}^{+0.1}$ &... 	  	 \\
Q1211p1030$\_$38$^a$	&0.6296	&20.30$\pm$0.24	&-2.82$^{+0.89}_{-0.89}$	&0.63031  	&138.4	&0.71 $\pm$0.26	   	&10.5$_{-0.5}^{+0.5}$ &8.44$\pm$0.1	 \\
Q1211p1030$\_$57$^a$	&0.6296	&20.30$\pm$0.24	&-2.82$^{+0.89}_{-0.89}$	&0.62957  	&16.9	$<$&0.386		&11.4$_{-0.4}^{+0.3}$ &... 	  	 \\
Q1211p1030$\_$13$^a$	&0.8999	&$<$18.5	&$<$-1.2			&0.89909  	&175.3	&0.46 $\pm$0.19	   	& 8.9$_{-0.7}^{+0.7}$ &8.63$\pm$0.2	 \\
Q1211p1030$\_$16$^a$	&0.8999	&$<$18.5	&$<$-1.2			&0.8994	  	&77.4	&0.91 $\pm$0.36	   	&                 ... &... 	  	 \\
Q1211p1030$\_$48	&0.8999	&$<$18.5	&$<$-1.2			&0.9	  	&52.8	&3.49 $\pm$0.31	   	&                 ... &8.28$\pm$0.05	 \\
Q1211p1030$\_$58$^a$	&0.8999	&$<$18.5	&$<$-1.2			&0.89894  	&268.6	&0.08 $\pm$0.36	   	&                 ... &8.46$\pm$0.1	 \\
Q1211p1030		&1.0496	&$<$18.90	&$<$-1.3			&... 	     	&...  	&...  	        	&		  ... &... 	  	 \\
Q1229m021$\_$39$^c$	&0.3950	&20.75$\pm$0.07	&-0.18$^{+0.14}_{-0.14}$	&0.39531  	&6.8	&0.89 $\pm$0.49	   	& 9.5$_{-0.2}^{+0.3}$ &8.63$\pm$0.16	 \\
Q1229m021$\_$13$^c$	&0.7572	&18.36$\pm$0.09	&-0.2$^{+0.1}_{-0.1}$		&0.7575	  	&70.9	&2.18 $\pm$0.75	   	&                 ... &8.42$\pm$0.14	 \\
Q1229m021$\_$5		&0.7691	&18.11$\pm$0.15	&-3.1$^{+0.2}_{-0.1}$		&0.76855  	&178.3	&1.66 $\pm$0.61	   	& 9.8$_{-0.7}^{+0.7}$ &8.3 $\pm$0.14	 \\
Q1229m021$\_$29		&0.7691	&18.11$\pm$0.15	&-3.1$^{+0.2}_{-0.1}$		&0.76852  	&135.9	&0.12 $\pm$0.51	   	& 9.6$_{-0.7}^{+0.7}$ &8.6 $\pm$0.11	 \\
Q1229m021$\_$40		&0.7691	&18.11$\pm$0.15	&-3.1$^{+0.2}_{-0.1}$		&0.76646  	&60.0	&3.0  $\pm$0.95	   	& 9.7$_{-0.7}^{+0.7}$ &8.49$\pm$0.1	 \\
Q1229m021$\_$6$^a$	&0.8311	&18.84$\pm$0.10	&-2.6$^{+0.1}_{-0.1}$		&0.8317	  	&182.3	&0.51 $\pm$0.21	   	&                 ... &8.34$\pm$0.19	 \\
Q1229m021$\_$8$^a$	&0.8311	&18.84$\pm$0.10	&-2.6$^{+0.1}_{-0.1}$		&0.8306	  	&125.5	&$<$0.333		& 8.5$_{-0.6}^{+0.7}$ &... 	  	 \\
Q1229m021$\_$10$^a$	&0.8311	&18.84$\pm$0.10	&-2.6$^{+0.1}_{-0.1}$		&0.8319	  	&173.6	&6.23 $\pm$2.13	   	& 9.7$_{-0.7}^{+0.7}$ &... 	  	 \\
Q1229m021$\_$41$^a$	&0.8311	&18.84$\pm$0.10	&-2.6$^{+0.1}_{-0.1}$		&0.83199  	&186.8	&7.79 $\pm$11.62	&                 ... &... 	  	 \\
Q1342m0035$\_$4$^a$	&0.5380	&19.78$\pm$0.13	&-1.2$^{+0.1}_{-0.1}$		&0.53773  	&43.6	&3.9  $\pm$1.21	   	&10.1$_{-0.1}^{+0.1}$ &8.67$\pm$0.06	 \\
Q1342m0035$\_$9$^a$	&0.5380	&19.78$\pm$0.13	&-1.2$^{+0.1}_{-0.1}$		&0.53877  	&24.3	&0.44 $\pm$0.18	   	&10.9$_{-0.1}^{+0.2}$ &8.6 $\pm$0.18	 \\
Q1345m0023$\_$13$^a$	&0.6057	&18.85$\pm$0.20	&-0.9$^{+0.2}_{-0.2}$		&0.60697  	&55.6	&3.06 $\pm$1.04	   	&10.1$_{-0.6}^{+0.6}$ &8.64$\pm$0.07	 \\
Q1345m0023$\_$40$^a$	&0.6057	&18.85$\pm$0.20	&-0.9$^{+0.2}_{-0.2}$		&0.6054	  	&162.4	&0.51 $\pm$0.18	   	& 8.7$_{-0.6}^{+0.6}$ &8.31$\pm$0.14	 \\
Q1431m0050$\_$10$^a$	&0.6085	&19.18$\pm$0.24	&-0.8$^{+0.1}_{-0.1}$		&0.60843  	&45.4	&14.27$\pm$4.44	   	& 9.8$_{-0.1}^{+0.1}$ &8.62$\pm$0.05	 \\
Q1431m0050$\_$68$^a$	&0.6085	&19.18$\pm$0.24	&-0.8$^{+0.1}_{-0.1}$		&0.60742  	&73.0	&1.84 $\pm$0.72	   	&10.6$_{-0.3}^{+0.2}$ &... 	  	 \\
Q1431m0050$\_$26$^a$	&0.6868	&18.40$\pm$0.07	&-1.2$^{+0.029}_{-0.012}$	&0.6871	  	&23.3	&0.42 $\pm$0.18	   	&                 ... &... 	  	 \\
Q1431m0050$\_$73$^a$	&0.6868	&18.40$\pm$0.07	&-1.2$^{+0.029}_{-0.012}$	&0.68687  	&188.7	&0.89 $\pm$0.39	   	& 8.9$_{-0.6}^{+0.6}$ &... 	  	 \\
Q1515p0410$\_$4$^a$	&0.5592	&20.20$\pm$0.19	&0.20$^{+0.24}_{-0.24}$		&0.55859  	&10.1	&$<$0.171		&11.0$_{-0.2}^{+0.1}$ &... 	  	 \\
Q1515p0410$\_$9$^a$	&0.5592	&20.20$\pm$0.19	&0.20$^{+0.24}_{-0.24}$		&0.55866  	&221.0	&$<$0.171		&10.2$_{-0.1}^{+0.2}$ &8.75$\pm$0.08	 \\
Q1515p0410$\_$11	&0.5592	&20.20$\pm$0.19	&0.20$^{+0.24}_{-0.24}$		&0.55995  	&216.0	&$<$0.171		&10.1$_{-0.1}^{+0.2}$ &... 	  	 \\
Q1515p0410$\_$13$^a$	&0.5592	&20.20$\pm$0.19	&0.20$^{+0.24}_{-0.24}$		&0.5586	  	&194.1	&$<$0.171		&10.2$_{-0.1}^{+0.2}$ &... 	  	 \\
Q1515p0410$\_$42	&0.5592	&20.20$\pm$0.19	&0.20$^{+0.24}_{-0.24}$		&0.55743  	&171.3	&$<$0.171		& 9.1$_{-0.6}^{+0.6}$ &8.53$\pm$0.15	 \\
Q1515p0410$\_$54$^a$	&0.5592	&20.20$\pm$0.19	&0.20$^{+0.24}_{-0.24}$		&0.55892  	&168.8	&$<$0.171		&                 ... &8.61$\pm$0.11	 \\
Q1554m203$\_$51$^a$	&0.7869	&$<$19.00	&$<$-1.6			&0.78558  	&23.0	&0.56 $\pm$0.2		& 9.4$_{-0.5}^{+0.5}$ &... 	  	 \\
Q2131m1207$\_$5$^a$	&0.4298	&19.50$\pm$0.15	&-1.47$^{+0.22}_{-0.22}$	&0.42999  	&48.5	&0.11 $\pm$0.05	   	&10.1$_{-0.1}^{+0.2}$ &8.43$\pm$0.12	 \\
Q2131m1207$\_$26$^a$	&0.4298	&19.50$\pm$0.15	&-1.47$^{+0.22}_{-0.22}$	&0.42971  	&172.1	&0.24 $\pm$0.09	   	&                 ... &8.12$\pm$0.11	 \\
Q2131m1207$\_$34$^a$	&0.4298	&19.50$\pm$0.15	&-1.47$^{+0.22}_{-0.22}$	&0.4308	  	&59.2	&0.05 $\pm$0.03	   	&                 ... &... 	  	 \\
Q2131m1207$\_$43$^a$	&0.4298	&19.50$\pm$0.15	&-1.47$^{+0.22}_{-0.22}$	&0.43003  	&145.9	&2.47 $\pm$0.82    	&                 ... &8.68$\pm$0.05     \\
Q2353m0028   	      	&0.6044 &21.54$\pm$0.15	&-1.18$^{+0.41}_{-0.41}$	&... 	     	&...  	&...  	        	&		  ... &... 	  	 \\	  		
\hline\hline 				       			 	 
\end{longtable}
\tablefoot{ This table summarises for each of the quasar fields, the main physical properties of the atomic gas probed in absorption (redshift, column density, neutral gas metallicity) and the galaxies associated with the strong \hi\ quasar absorbers (redshift, distance from quasar sightline, star formation rate, stellar mass and ionised gas metallicity). The former absorption properties are estimated from HST and VLT/Keck spectroscopy of the background quasar while the latter are derived from VLT/MUSE and HST imaging emission observations combined. Entries with no number underscored refer to the 5 quasar absorbers with no associated galaxies identified in MUSE fields. \\
{\bf References}:
  $^1$ \cite{Peroux22}; $^2$ \cite{Halley26};  $^3$ \cite{Weng23a};  $^4$ \cite{Augustin24}.\\
  $^a$ denotes ALMA Large Programme targets (this work);
  $^b$ \cite{Kanekar18};
  $^c$ \cite{Szakacs21};
  $^d$ \cite{Klitsch18};
  $^e$ \cite{Peroux19}.\\
  }

\end{appendix}

\end{document}